\documentclass[twocolumn, onlinenumbers, hyperref={colorlinks,linkcolor=blue,citecolor=blue,urlcolor=red}]{aastex}

\usepackage{threeparttable}
\usepackage{graphicx}
\usepackage{placeins}
\usepackage{float}
\usepackage{url}
\usepackage{natbib}
\usepackage{xspace}
\usepackage{amssymb}

\def\kms{\,{\rm {km\, s^{-1}}}}

\def\Msun{{\rm M_\odot}}

\def\gs{\mathrel{\raise1.16pt\hbox{$>$}\kern-7.0pt
\lower3.06pt\hbox{{$\scriptstyle \sim$}}}}
\def\ls{\mathrel{\raise1.16pt\hbox{$<$}\kern-7.0pt
\lower3.06pt\hbox{{$\scriptstyle \sim$}}}}
\def\gtsima{\, {\buildrel > \over \sim} \,}
\def\ltsima{\, {\buildrel < \over \sim} \,}
\def\prosima{\, {\buildrel \propto \over \sim} \,}
\def\gsim{\lower.5ex\hbox{\gtsima}}
\def\lsim{\lower.5ex\hbox{\ltsima}}
\def\simgt{\lower.5ex\hbox{\gtsima}}
\def\simlt{\lower.5ex\hbox{\ltsima}}
\def\simpr{\lower.5ex\hbox{\prosima}}

\begin{document}

\title{Halo mass-observable proxy scaling relations and their dependencies on galaxy and group properties}

\author[0000-0002-9272-5978]{Ziwen Zhang}
\affiliation{CAS Key Laboratory for Research in Galaxies and Cosmology, Department of Astronomy, University of Science and Technology of China, Hefei, Anhui 230026, China; Email: ziwen@mail.ustc.edu.cn, whywang@ustc.edu.cn}
\affiliation{School of Astronomy and Space Science, University of Science and Technology of China, Hefei 230026, China}

\author[0000-0002-4911-6990]{Huiyuan Wang}
\affiliation{CAS Key Laboratory for Research in Galaxies and Cosmology, Department of Astronomy, University of Science and Technology of China, Hefei, Anhui 230026, China; Email: ziwen@mail.ustc.edu.cn, whywang@ustc.edu.cn}
\affiliation{School of Astronomy and Space Science, University of Science and Technology of China, Hefei 230026, China}

\author[0000-0003-1297-6142]{Wentao Luo}
\affiliation{CAS Key Laboratory for Research in Galaxies and Cosmology, Department of Astronomy, University of Science and Technology of China, Hefei, Anhui 230026, China; Email: ziwen@mail.ustc.edu.cn, whywang@ustc.edu.cn}
\affiliation{School of Astronomy and Space Science, University of Science and Technology of China, Hefei 230026, China}

\author[0000-0001-5356-2419]{Houjun Mo}
\affiliation{Department of Astronomy, University of Massachusetts, Amherst, MA 01003-9305, USA}

\author[0000-0003-0002-630X]{Jun Zhang}
\affiliation{Department of Astronomy, Shanghai Jiao Tong University, Shanghai 200240, China}
\affiliation{Shanghai Key Laboratory for Particle Physics and Cosmology, Shanghai 200240, China}

\author{Xiaohu Yang}
\affiliation{Department of Astronomy, Shanghai Jiao Tong University, Shanghai 200240, China}
\affiliation{Shanghai Key Laboratory for Particle Physics and Cosmology, Shanghai 200240, China}

\author{Hao Li}
\affiliation{CAS Key Laboratory for Research in Galaxies and Cosmology, Department of Astronomy, University of Science and Technology of China, Hefei, Anhui 230026, China; Email: ziwen@mail.ustc.edu.cn, whywang@ustc.edu.cn}
\affiliation{School of Astronomy and Space Science, University of Science and Technology of China, Hefei 230026, China}

\author[0000-0003-3616-6486]{Qinxun Li}
\affiliation{CAS Key Laboratory for Research in Galaxies and Cosmology, Department of Astronomy, University of Science and Technology of China, Hefei, Anhui 230026, China; Email: ziwen@mail.ustc.edu.cn, whywang@ustc.edu.cn}
\affiliation{School of Astronomy and Space Science, University of Science and Technology of China, Hefei 230026, China}

\begin{abstract}
Based on the DECaLS shear catalog, we study the scaling relations between halo mass($M_{\rm h}$) and various proxies for SDSS central galaxies, including stellar mass($M_*$), stellar velocity dispersion($\sigma_*$), abundance matching halo mass($M_{\rm AM}$) and satellite velocity dispersion($\sigma_{\rm s}$), and their dependencies on galaxy and group properties. In general, these proxies all have strong positive correlations with $M_{\rm h}$, consistent with previous studies. 
We find that the $M_{\rm h}$-$M_*$ and $M_{\rm h}$-$\sigma_*$ relations depend strongly on group richness($N_{\rm sat}$), while the $M_{\rm h}$-$M_{\rm AM}$ and $M_{\rm h}$-$\sigma_{\rm s}$ relations are independent of it. Moreover, the dependence on star formation rate(SFR) is rather weak in the $M_{\rm h}$-$\sigma_*$ and $M_{\rm h}$-$\sigma_{\rm s}$ relations, but very prominent in the other two. 
$\sigma_{\rm s}$ is thus the best proxy among them, and its scaling relation is in good agreement with hydro-dynamical simulations. However, estimating $\sigma_{\rm s}$ accurately for individual groups/clusters is 
challenging because of interlopers and the requirement for sufficient satellites.
We construct new proxies by combining $M_*$, $\sigma_*$, and $M_{\rm AM}$, and find the proxy with 30\% contribution from $M_{\rm AM}$ and 70\% from $\sigma_*$ can minimize the dependence on $N_{\rm sat}$ and SFR. 
We obtain the $M_{\rm h}$-supermassive black hole(SMBH) mass relation via the SMBH scaling relation and find indications for rapid and linear growth phases for SMBH.
We also find that correlations among $M_{\rm h}$, $M_*$ and $\sigma_*$ change with $M_*$, indicating that different processes drive the growth of galaxies and SMBH at different stages. 
\end{abstract}


\section{Introduction} \label{sec:intro}

In the standard $\Lambda$CDM paradigm, dark matter halos, approximately in 
dynamical equilibrium, form hierarchically through gravitational instability. 
Galaxies form and evolve in the potential wells of dark matter halos 
\citep[e.g.][]{WhiteRees1978, FallEfstathiou1980, Mo2010}. Galaxies residing in one
halo are regarded as a galaxy group or a cluster. 
In recent years, many studies have 
concentrated on establishing galaxy - halo connections, manifested as scaling 
relations between halo mass and galaxy and group properties (proxies). 
These scaling relations can be used to estimate the halo mass of galaxy 
groups and clusters \citep{Yang2005, Zahid2016, Seo2020},
constrain cosmology \citep{Bocquet2015, Caldwell2016},  
and study galaxy formation models \citep[see][for references]{Mo2010,Wechsler2018,Chen2021a,Chen2021b}.

In these scaling relations, halo mass ($M_{\rm h}$) is not a direct observable. $M_{\rm h}$ 
estimation usually requires high-quality data and/or sophisticated techniques, 
and sometimes model assumptions. Numerous studies have developed various methods 
to constrain the halo mass, including the virial mass estimator 
\citep[e.g.][]{Carlberg2001, Rines2013, Abdullah2020}, X-ray emission 
\citep{Ruel2014}, Sunyaev–Zel’dovich (SZ) effects \citep{Ruel2014, Rines2016}, caustics technique \citep[e.g.][]{Rines2006, Rines2013}, abundance matching \citep{Yang2007} and weak lensing \citep[e.g.][]{Mandelbaum2006, Mandelbaum2016, Hoekstra2007, Leauthaud2012, Velander2014MNRAS.437.2111V, Gonzalez2015, Han2015, Hudson2015, Viola2015MNRAS.452.3529V, Luo2018ApJ, Gonzalez2021, Rana2022, Zhang2021, Zhang2022}. Various halo mass proxies 
have been investigated, such as stellar mass, group 
richness, stellar velocity dispersion, group total stellar mass or luminosity, and satellite velocity dispersion.

Among these studies, the most widely discussed halo mass proxy is the stellar mass ($M_*$) of central galaxies. And the corresponding scaling relation is referred to as stellar mass-halo mass relation (hereafter SHMR) \citep[e.g.][]{Yang2003, Moster2010, Leauthaud2012, Li2013, Mandelbaum2016, Kravtsov2018, Wechsler2018, Behroozi2019, Zhang2021,Zhang2022}. The SHMR can be used to constrain galaxy formation 
processes. For example, $M_*/M_{\rm h}/(\Omega_{\rm b}/\Omega_{\rm m})$ is usually regarded as 
the baryon-to-star conversion efficiency. The mean efficiency peaks around the Milky-way like galaxies, with a value of about 20\%, and declines quickly toward both lower and higher $M_*$ ends. At lower $M_*$, the efficiency is believed to be suppressed by supernova feedback and stellar winds \citep[e.g.][]{Dekel1986, Kauffmann1998MNRAS.294..705K, Cole2000MNRAS.319..168C, Hopkins2012}, while the feedback from active galactic nuclei (AGN) is suggested to be responsible for suppressing the efficiency at higher $M_*$ 
\citep[e.g.][]{Silk1998, Croton-06, Fabian-12, Heckman2014, CuiW2021}. 
The SHMR has large scatter, which depends on galaxy properties, such as color, star 
formation rate and morphology \citep[e.g.][]{More2011,Rodriguez-Puebla2015ApJ...799..130R, Mandelbaum2016, Behroozi2019, Lange2019, Romeo2020, Zhang2021, Bilicki2021, Posti2021, Xu2022, Wangk2023a}. 
Research based on simulations suggests that the dependence is 
related to both halo assembly and AGN feedback \citep{CuiW2021}. 
More recently, several studies found a number of massive star-forming/spiral galaxies 
that exhibit very high star formation efficiency, about 50\%-60\% 
\citep[e.g.][]{Posti2019, Zhang2022}. Their results indicate that these galaxies 
can convert most of their halo gas into stars, implying that AGN feedback is 
inefficient in them. The scatter of SHMR depends also strongly on satellite 
properties \citep[][]{Lu2016, Zhou2022}. 
All these results suggest that the SHMR and its scatter contain valuable 
information about galaxy formation and halo assembly.

Stellar velocity dispersion ($\sigma_*$) directly reflects the 
gravitational potential of the galaxy, and is well correlated with the 
stellar mass. Given the SHMR, one may expect a strong correlation between 
stellar velocity dispersion and halo mass. The $\sigma_*$ measurement is 
straightforward, and its systematic uncertainty may be lower than the stellar mass, 
making the stellar velocity dispersion an ideal alternative to stellar mass 
as a halo-mass proxy. Previous studies on this connection find that the 
stellar velocity dispersion is indeed a robust proxy for halo mass 
\citep[e.g.][]{ Zahid2016, Zahid2018, Seo2020, Sohn2020}. 
The $M_{\rm h}$-$\sigma_*$ relation also has other important applications.
For instance, \cite{Shankar2020} found that the $M_{\rm h}$-$\sigma_*$ relation 
and the large-scale clustering of AGNs can be used to constrain the shape and 
amplitude of the scaling relation between supermassive black hole (SMBH) mass and galaxy properties, 
as well as to constrain the radiative efficiency of AGNs. 
The halo masses in these observational studies are estimated via abundance 
matching and satellite kinematics. It is required to have a weak lensing based investigation as an independent check. Moreover, it is also interesting to 
investigate if the SHMR and $M_{\rm h}$-$\sigma_*$ relation are independent of each other.
Studying the bivariate correlation may be particularly important for 
understanding the connection between AGN feedback, galaxy growth and halo environments.

Previous studies have also adopted the total stellar mass or total luminosity 
of group members as a halo mass proxy, and found good correlations with halo 
mass \citep[e.g.][]{Yang2005,Conroy07apj, Han2015}. \cite{Han2015} performed a 
maximum-likelihood weak-lensing analysis and found that the group luminosity 
has the tightest relation with weak lensing halo mass among all single proxies 
they investigated. The group luminosity and stellar mass have been used to 
estimate the halo mass based on abundance matching \citep[e.g.][]{Yang2005,Yang2007}. 
The inferred halo mass is referred to as the abundance matching halo mass, 
denoted by $M_{\rm AM}$, in the following.
\cite{Luo2018ApJ} and \cite{Gonzalez2021} studied the relation between $M_{\rm AM}$ 
and halo mass based on weak lensing. \cite{Gonzalez2021} further investigated the 
dependence of the $M_{\rm h}$-$M_{\rm AM}$ relation on galaxy morphology 
and found that the relation for early-type centrals is closer to the 
one-to-one relation than the relation based on all types of centrals. 

Satellite kinematics, which results from the equilibrium of a gravitational
system, can also provide a measurement of halo mass.    
Many investigations have been carried out to study
the relation between satellite velocity dispersion ($\sigma_{\rm s}$) and halo mass 
using both simulations \citep[e.g.][]{Evrard2008, Li2012, Munari2013, Saro2013} and 
observational data \citep[e.g.][]{Rines2006, Yang2007, Hoekstra2007, Rines2013, Ruel2014, Gonzalez2015, Han2015, Viola2015MNRAS.452.3529V, Rines2016, Abdullah2020, Gonzalez2021, Rana2022}.
\cite{Evrard2008} used dark matter particles to calculate $\sigma_{\rm s}$ and found that the 
relation has small scatter. \cite{Munari2013} and \cite{Saro2013} found 
a weak bias (less than 10$\%$) when using simulated subhalos or satellite 
galaxies to derive $\sigma_{\rm s}$. They all found that halo mass is roughly 
proportional to $\sigma_{\rm s}^3$, consistent with theoretical expectations.

Observationally, various methods have been applied to establish the 
$M_{\rm h}$-$\sigma_{\rm s}$ relation. For example, some studies estimated the 
halo mass based on the caustics technique \citep[e.g.][]{Rines2006,Rines2013}, 
SZ effect \citep[e.g.][]{Ruel2014, Rines2016} and X-ray observation \citep{Ruel2014}. These
studies usually focused on individual galaxy clusters and the obtained relations 
are broadly consistent with those from simulations. However, the results based 
on weak lensing data appear complex. 
Some studies obtained similar slopes but different amplitudes in comparison 
to relations obtained from simulations \citep[e.g.][]{Gonzalez2021, Zhang2022}, 
and some obtained even different slopes \citep[e.g.][]{Han2015, Gonzalez2015, 
Viola2015MNRAS.452.3529V, Rana2022}. \cite{Hoekstra2007} obtained the relation 
for massive clusters and found it in agreement with simulations but with 
larger scatter. In addition, very few studies extended the 
$M_{\rm h}$ - $\sigma_{\rm s}$ relation to $M_{\rm h} < 10^{13}\rm M_{\odot}$
\citep[e.g.][]{Han2015, Zhang2022}.

The main goal of this paper is to study various halo mass proxies 
(stellar mass $M_*$, stellar velocity dispersion $\sigma_*$, abundance matching halo mass $M_{\rm AM}$, 
satellite velocity dispersion $\sigma_{\rm s}$, and combinations of them) and their 
relations with the halo mass obtained from weak lensing. We also investigate 
their dependencies on other galaxy/group properties to shed light 
on underlying physical processes, and evaluate the performance of these 
halo mass proxies. 
The outline of this paper is as follows. In Section \ref{sec_data}, we describe sample selections and the weak lensing shear catalog to be used. 
Section \ref{sec_measurements} presents measurements of weak lensing and 
satellite kinematics.  We also introduce in this section an proxy 
based on combinations of parameters and the way to evaluate the performance 
of halo mass proxies. In Section \ref{sec_scaling_relations}, 
we show the scaling relations we obtain and their dependencies on 
other galaxy and group properties. Finally, we summarize our 
results in Section \ref{sec_summary}. Throughout this paper, 
we assume a flat $\Lambda$CDM cosmology with $\rm \Omega_m$ = 0.307, 
$\rm \Omega_b$ = 0.048, $\rm \Omega_\Lambda$ = 0.693 and h = 0.678. 
Here, h = $\rm H_0/100$ km $\rm s^{-1}$ $\rm Mpc^{-1}$ 
and $\rm H_0$ is the Hubble constant \citep{Planck2016}.

\section{Observational data}\label{sec_data}
\subsection{Central galaxy samples}\label{sec_galaxy_samples}

The galaxies used in this paper are drawn from the New York University Value Added Galaxy Catalog 
\citep[NYU-VAGC\footnote{http://sdss.physics.nyu.edu/vagc/}][]{Blanton-05a} of the Sloan 
Digital Sky Survey Data Release 7 (SDSS DR7) \citep{Abazajian-09}. We select galaxies with 
$r$-band Petrosian magnitude $r\le 17.72$, with redshift in the range of 0.01 $\le$ $z$ $\le$ 0.2 
and with redshift completeness $C_z >$ 0.7. 
We use the group catalog\footnote{ https://gax.sjtu.edu.cn/data/Group.html} of \cite{Yang2005, Yang2007} to classify central and satellite galaxies.
Central galaxies are defined as the most massive galaxies in galaxy groups.
As shown in the next section, we obtain the halo mass by using the Dark Energy Camera Legacy Survey \citep[DECaLS,][]{Dey2019AJ}. About 32$\%$ of SDSS DR7 galaxies do not overlap with the DECaLS. This leaves a sample of 323,448 central galaxies. 

$M_*$ of the central galaxies are obtained 
by cross-matching with the MPA-JHU DR7 catalog \footnote{https://wwwmpa.mpa-garching.mpg.de/SDSS/DR7/} \citep{Kauffmann2003, Brinchmann2004}. They are calculated by fitting the SDSS $ugriz$ photometry to models of galaxy spectral energy distribution.
$\sigma_*$ is provided by NYU-VAGC. 
$M_{\rm AM}$ for each central galaxy is provided by the group catalog\citep{Yang2007}. Here, we mainly use $M_{\rm AM}$ based on the group total stellar mass.
The estimate of $\sigma_{\rm s}$ will be described in Section \ref{sec_sk} and the combined proxies are introduced in Section \ref{sec_new_proxy}.

Not all galaxies have valid estimations of $M_*$, $\sigma_*$, star-formation rate (SFR) and $M_{\rm AM}$. For example, a small fraction of galaxies have no measurement of $\sigma_*$ (usually low mass galaxies), and some have unreasonably high $\sigma_*$ (e.g. $\sigma_*\sim 1000\kms$). Moreover, a fraction of low mass galaxies have no
available $M_{\rm AM}$ because $M_{\rm AM}$ can only be estimated in a complete sample. 
We thus construct two total samples. The first one, referred to as ATotal, has 
measurements of $M_*$ and SFR and $0<\sigma_*/\kms<630$ of 299,806 galaxies. 
This sample is used to study the $M_{\rm h}$-$M_*$ and $M_{\rm h}$-$\sigma_*$ relations (Section \ref{sec_SHMR} and \ref{sec_Mh_g_sigma}).  The second one, referred to as 
BTotal, requires measurements of $M_{\rm AM}$ in addition. This requiement 
removes 60,769 galaxies. The BTotal sample is used to study the 
$M_{\rm h}$-$M_{\rm AM}$ relation (Section \ref{sec_MAM}). When we study 
the $M_{\rm h}$-$\sigma_{\rm s}$ relation (Section \ref{sec_Mh_sigmas}), 
both samples are used. We do not use BTotal sample
to study the $M_{\rm h}$-$M_*$ and $M_{\rm h}$-$\sigma_*$ relations, 
because this sample excludes many low-mass galaxies. We list the sample 
selection, the sample size and the sections where the samples are 
used in Table \ref{tab_samples}.

We will also consider the dependencies of the various 
relations mentioned above on several galaxy properties, such as $M_*$, 
$\sigma_*$ and SFR, and one group property,  the 
group richness. The galaxy SFR is obtained by cross-matching with the MPA-JHU DR7 
catalog, where the SFRs of individual galaxies 
were derived from spectroscopic and photometric data of the SDSS. 
We adopt the demarcation line proposed by \cite{Bluck2016}, to separate star-forming and quenched galaxies.
Thus, the ATotal (BTotal) sample is divided into ASF (BSF) and AQ (BQ) subsamples,
corresponding to star-forming and quenched galaxies, respectively. 
For each central galaxy, its group richness is quantified in terms of
the number of satellites (hereafter $N_{\rm sat}$) provided by the group catalog. 
We divide the ATotal (BTotal) sample into four subsamples, AC0, AC1, AC2 and AC3 
(BC0, BC1, BC2 and BC3), where AC0 and BC0 have $N_{\rm sat}=0$, 
AC1 and BC1 have $N_{\rm sat}\geq 1$, AC2 and BC2 have $1\leq N_{\rm sat}\leq 2$, 
and AC3 and BC3 have $N_{\rm sat}\geq 3$. 
These subsamples are also listed in Table \ref{tab_samples}. These samples
are sometimes referred to as the A-series and B-series samples, respectively.

\begin{table*}
\centering
\caption{Sample selection used in our analysis}
\label{tab_samples}
\begin{threeparttable}
\setlength{\tabcolsep}{6.5mm}{
\begin{tabular}{c c c c c}
\hline
Sample\tnote{(a)} & Sample criteria\tnote{(b)} &  subsample criteria\tnote{(c)} & 
Sample Size &  Section\tnote{(d)}\\
\hline
ATotal &       $0<\sigma_*<630$    &       & 299,806 & \ref{sec_sk}, \ref{sec_SHMR}, \ref{sec_Mh_g_sigma}, \ref{sec_Mh_sigmas} \\
AC0  & available $M_*$  & $N_{\rm sat} = 0$            & 262,404 & \ref{sec_sk}, \ref{sec_SHMR}, \ref{sec_Mh_g_sigma} \\
AC1   &  available SFR    & $N_{\rm sat} \geq 1$         & 37,402 & \ref{sec_sk}, \ref{sec_SHMR}, \ref{sec_Mh_sigmas} \\
AC2   &      & $1 \leq N_{\rm sat} \leq 2$  & 30,245 & \ref{sec_sk}, \ref{sec_SHMR}, \ref{sec_Mh_g_sigma}  \\
AC3   &      & $N_{\rm sat} \geq 3$         & 7,157 & \ref{sec_sk}, \ref{sec_SHMR}, \ref{sec_Mh_g_sigma}  \\
ASF   &      & star-forming         & 176,202 & \ref{sec_SHMR}, \ref{sec_Mh_g_sigma}   \\
AQ   &      & quenched         & 123,604 & \ref{sec_SHMR}, \ref{sec_Mh_g_sigma}   \\
\hline
BTotal &       $0<\sigma_*<630$      &       & 239,037 & \ref{sec_MAM}, \ref{sec_combined_proxy}  \\
BC0   &  available $M_*$        & $N_{\rm sat} = 0$            & 204,514 & \ref{sec_MAM}, \ref{sec_combined_proxy} \\
BC1   &  available SFR             & $N_{\rm sat} \geq 1$         & 34,523  & \ref{sec_Mh_sigmas} \\
BC2   &  available $M_{\rm AM}$  & $1 \leq N_{\rm sat} \leq 2$  & 27,477  & \ref{sec_MAM}, \ref{sec_Mh_sigmas} \\
BC3   &     & $N_{\rm sat} \geq 3$         & 7,046  & \ref{sec_MAM}, \ref{sec_Mh_sigmas}, \ref{sec_combined_proxy}  \\
BSF   &      & star-forming         & 123,796 & \ref{sec_MAM}, \ref{sec_Mh_sigmas}, \ref{sec_combined_proxy}   \\
BQ   &      & quenched         & 115,241 & \ref{sec_MAM}, \ref{sec_Mh_sigmas}, \ref{sec_combined_proxy}   \\
\hline
\end{tabular}}

\begin{tablenotes}
\footnotesize
\item[(a)] ATotal and BTotal are two total samples, and the rest are corresponding subsamples.
\item[(b)] The selection criteria is applied to the corresponding total samples and subsamples.

\item[(c)] The selection criterion is only applied to the corresponding subsample.
\item[(d)] The sections where the samples are used.
\end{tablenotes}
\end{threeparttable}
\end{table*} 

\subsection{DECaLS shear catalog}

We use the shear catalog based on the DECaLS DR8 imaging data \citep{Dey2019AJ,Zou_2019} 
to measure galaxy-galaxy lensing signals. 
Galaxy shapes are measured by using the FOURIER$\_$QUAD pipeline
which measures galaxy shear with great accuracy even for 
extremely faint images (signal-to-noise ratio $<$ 10). The pipeline 
was tested both with simulations \citep{Zhang2015JCAP} and with 
observations (See \cite{Zhang_2019} for the CFHTLenS data and 
\cite{Wanghr2021, ZhangJ2022} for the DECaLS data).
The shear catalog covers more than ten thousand square degrees 
in the $g$, $r$, and $z$ bands, with 99, 111 and 116 million distinct 
galaxies, respectively. The FOURIER$\_$QUAD method counts images of 
the same galaxy but in different exposures as different images. 

A machine learning algorithm based on decision trees is implemented to 
calculate photometric redshifts of galaxies in the shear catalog 
\citep{zhourongpu2021}. Eight parameters, including the $r$-band magnitude, 
($g - r$), ($r - z$), ($z - W1$), and ($W1 - W2$) colors, half-light radius, 
axial ratio, and shape probability, are adopted in the training progress. 
The photo-$z$ error of each individual shear image is obtained by 
perturbing the photometry of the galaxy. The procedures are as follows: 
(i) the uncertainty was assumed to follow a Gaussian distribution 
with the standard deviation equal to the photometric error; 
(ii) a random value generated from the distribution was added to the observed flux in 
each band to obtain a “perturbed” flux; 
(iii) the perturbation was repeated multiple times, and the standard deviation of the 
photo-$z$ estimates from the perturbations was used as the error of the 
photo-$z$ \citep[see details in][]{zhourongpu2021}.  

\section{Methods of analysis}\label{sec_measurements}

In this section, we describe the methods to measure halo mass and satellite kinematics from observational data. 
In addition, we present our methods for constructing  combined halo mass proxies and for evaluating the performance of these proxies.

\subsection{Weak lensing measurements}\label{sec_weak_lensing}

We measure the excess surface density (ESD) by using the $r$- and $z$-band data of the DECaLS shear catalog. 
\cite{Zhang2017} proposed a probability distribution function (PDF) symmetrization method to minimize the statistical uncertainty in shear signal. We apply a modified version of this method and estimate the ESD in physical coordinates by
\begin{equation}
    \Delta \Sigma(\rm R) = \gamma_t(\rm R)\Sigma_{\rm crit}.
\end{equation}
The details and general discussion about different sources of systematic errors of the PDF-symmetrization method in measuring the ESD signal are given in a companion paper \citep{Wangjq2022}. Error bars of ESD signals are estimated by using 150 bootstrap samples \citep{Barrow-Bhavsar-Sonoda-84}.

To model the ESD signal, we follow previous studies \citep[e.g.][]{Mandelbaum2008, Leauthaud2010,Luo2018ApJ, Zhang2022} and apply a halo model consisting of three terms:
\begin{equation}
    \Delta \Sigma = \Delta \Sigma_{\rm stellar} + \Delta \Sigma_{\rm NFW} + \Delta \Sigma_{\rm 2h}.
\end{equation}
The first term is the stellar mass term that represents the contribution from the galaxy stellar mass. We treat the galaxy as a point mass and adopt the average stellar mass of the galaxy sample directly from the observation. The second term is the one-halo term, which is the contribution from the host dark matter halo, and we assume the halo to follow the Navarro–Frenk–White \citep[NFW;][]{Navarro1997} density profile. The NFW profile has two free parameters: the dark halo mass($m_{\rm h}$) and the halo concentration($c$). The dark halo mass $m_{\rm h}$ is defined as the total dark matter mass within a spherical region of radius $r_{\rm 200m}$. Inside this region, the mean mass density is equal to 200 times the mean matter density of the Universe. We use the central galaxy as the tracer of the halo center. We adopt the equations from \cite{Yang2006} to analytically calculate the one-halo term. In the modeling process, we ignore the off-center effect \citep{Mandelbaum2016}, which has been shown to be a minor factor in halo mass estimation \citep{Wangjq2022}. 
The third term is the two-halo term, which represents the contribution from other halos. To calculate this term, we project the halo-matter cross-correlation function, $\xi_{\rm hm}$, along the line-of-sight. Here, $\xi_{\rm hm}$ = $b(m_{\rm h})\xi_{\rm mm}$, with $b(m_{\rm h})$ being the halo bias \citep[e.g.][]{Mo1996,Tinker2010} 
and $\xi_{\rm mm}$ being the linear matter-matter correlation function. 
We use {\it COLOSSUS} \citep{Diemer2018ApJS} to model both $b(m_{\rm h})$ and $\xi_{\rm mm}$.

There are two free parameters ($m_{\rm h}$ and $c$) in our model. The priors of these parameters are set to be flat, with $m_{\rm h}$ in the range of [10.0, 16.0] in logarithmic space and $c$ in the range of [1.0, 16.0].
To constrain them, we use $emcee$ \citep{Foreman-Mackey2013PASP} to run a Monte Carlo Markov Chain (MCMC).
In the following, the results shown are the median values of the posteriors, and the error bars correspond to the 16 and 84 percent of the posterior distribution (See \cite{Luo2018ApJ} for more details on the modeling and fitting process). Since we include the stellar mass term in our model, $m_{\rm h}$ only accounts for cold dark matter, diffuse gas, and satellites around centrals. Thus, throughout this paper, we use the 
total mass $M_{\rm h}=m_{\rm h} + M_{*}$ instead of $m_{\rm h}$ to represent the halo mass.

\subsection{Satellite kinematics} \label{sec_sk}

\begin{figure*}
    \centering
    \includegraphics[scale=0.11]{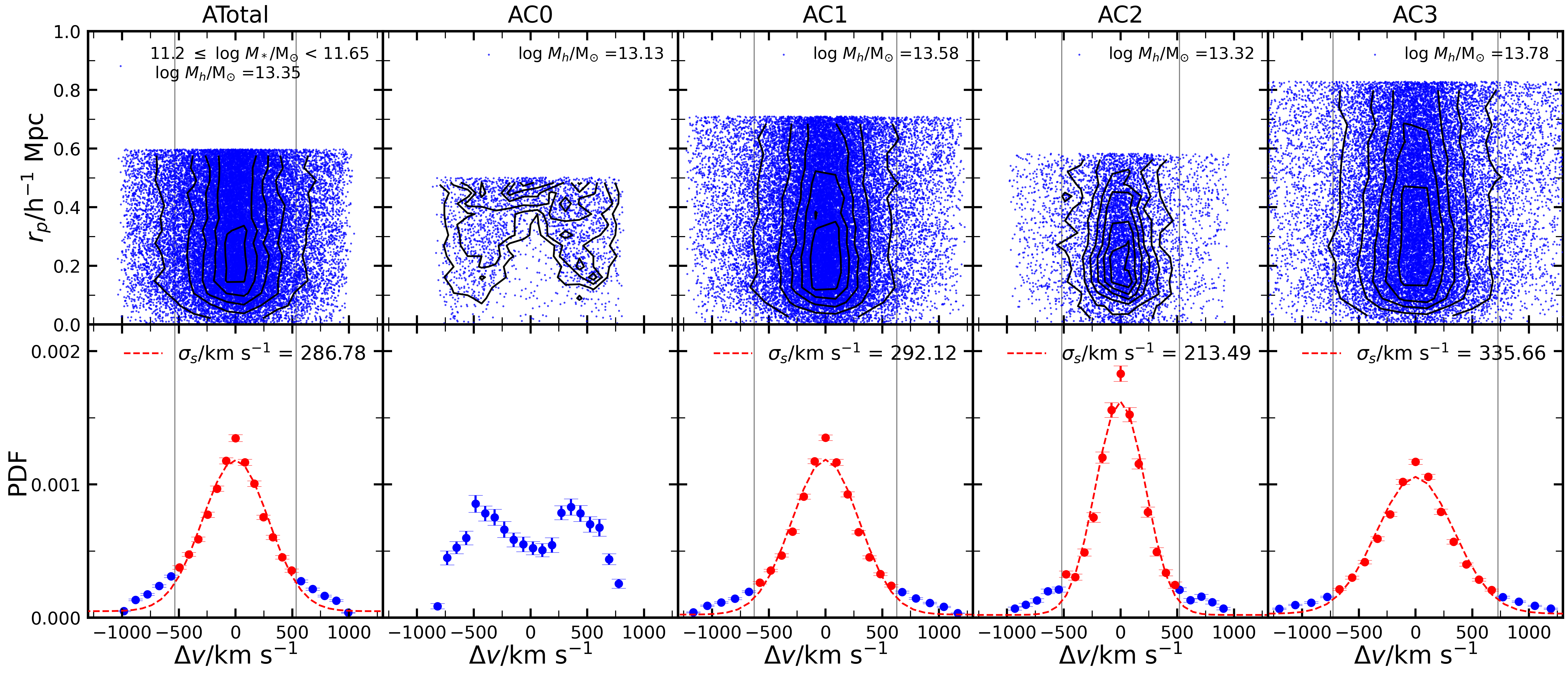}
    \caption{Upper panels: the data points and the contours show the distributions of satellite candidates in the projected phase space ($r_{\rm p}$-$\Delta v$). Lower panels: the probability distributions of $\Delta v$(red and blue symbols) and the best-fitting curves for PDFs within 1.5$v_{\rm 200m}$ (red symbols).  The two gray vertical lines in each panel correspond to $\Delta v = \pm1.5v_{\rm 200m}$. We only show the satellite candidates around centrals with $11.2\le\log M_*/\Msun< 11.65$ in ATotal, AC0, AC1, AC2 and AC3 samples. The halo masses and satellite velocity dispersion of the samples are listed in the upper and lower panels, respectively. Please see Section \ref{sec_sk} for details. }
    \label{fig_dv_rp}
\end{figure*}

Satellite kinematics can also serve as a powerful probe of dark matter halos
\citep[e.g.][]{McKay2002, vandenBosch2004, More2011, Wojtak2013, Lange2019a, Abdullah2020, Seo2020}. 
One major concern in satellite 
kinematics studies is the contamination of interlopers
\citep[see e.g.][]{vandenBosch2004, Mamon2010}. 
Recently, \cite{Zhang2022} used weak lensing technique to estimate the halo masses of their central galaxy samples and derived the corresponding halo virial radius, $r_{\rm 200m}$, and halo virial velocity, 
$v_{\rm 200m}$. They selected satellite candidates 
from a reference galaxy sample using the following set of criteria: 
$r_{\rm p}\le r_{\rm 200m}$, $|\Delta v|\le 3v_{\rm 200m}$, 
and $M_{\rm s}<M_*$. Here, $M_*$ and $M_{\rm s}$ are the stellar masses of 
the central and satellite candidate, respectively; 
$r_{\rm p}$ is the projected distance from the satellite candidate to its central;
and $\Delta v$ is the line-of-sight velocity difference between them. 
They found that the $M_{\rm h}$-$\sigma_{\rm s}$ relation so obtained 
has a power-law slope of about 3, but the amplitude of the relation is lower 
than that obtained from simulations (see Section \ref{sec_Mh_sigmas}). 

We apply the same method to select satellite candidates for central 
galaxies. The centrals are divided into samples based on either their 
$M_*$, or $\sigma_*$, or $M_{\rm AM}$ (see Section \ref{sec_Mh_sigmas}).
The distribution of the corresponding satellite candidates in the phase space ($r_{\rm p}$ versus $\Delta v$) are shown in Figure \ref{fig_dv_rp}. Here we only present the results for galaxies with $11.2\leq\log M_*/\Msun<11.65$ 
in A-series samples. 
The results for other galaxy samples/subsamples are similar. 
As one can see, there are cone-like structures in the phase space for ATotal, AC1, AC2 and AC3, as indicated by black contour lines, respectively.
The PDFs of $\Delta v$ are shown in the lower panels. The distributions for the four subsamples are very
similar to a Gaussian distribution. The peaks corresponding to the cone-like 
structures are contributed predominantly by satellites, while the tails that 
are outside of the cone are dominated by interlopers \citep{Mamon2010}. 
In contrast, AC0 galaxies has an empty cone-like structure and
double-peaked $\Delta v$ distribution. This is expected, 
because AC0 centrals have no satellite according to the group catalog, and 
the shown galaxies are mostly interlopers.

To eliminate interlopers as far as possible, we decide to only use satellite candidates with $|\Delta v| < |\Delta v|_{\rm cut}
=1.5v_{\rm 200m}$
to estimate the satellite velocity dispersion. Our visual inspection shows that the cut at $1.5v_{\rm 200m}$ 
roughly corresponds to the boundary of the cone-like structures, 
as shown by the gray vertical lines in Figure \ref{fig_dv_rp}.
We performed a series of tests using a value of $|\Delta v|_{\rm cut}$
ranging from 1.0 to 3.0$v_{\rm 200m}$, 
and found that the results are stable as long as $|\Delta v|_{\rm cut}$
is in the range of 1.3-1.8$v_{\rm 200m}$. A smaller $|\Delta v|_{\rm cut}$
excludes too many satellites in the cone-like structures and leads to poorer 
constraints. A larger $|\Delta v|_{\rm cut}$, 
on the other hand, leads to a larger dispersion because of the inclusion 
of more interlopers. Please see Section \ref{sec_Mh_sigmas} for more discussion.

The red symbols show the distributions of 
$|\Delta v|$ within $1.5v_{\rm 200m}$, with error bars 
estimated from 100 bootstrap samples. We fit the distributions to a functional form
that is a Gaussian plus a constant,
\begin{equation}\label{eq_fitting_sk}
P(\Delta v)=\frac{A}{\sqrt{2\pi}\sigma_{\rm v}} e^{-\Delta v^{2}/2\sigma^{2} _{\rm v}}+d. 
\end{equation}
The constant $d$ is used to account for interlopers within the cone.
We use the MCMC to fit the PDFs. The uncertainty of a SDSS galaxy redshift, 
which is about 35 km $\rm s^{-1}$, results in an error of 
$\sigma_{\rm e}$ = $\sqrt{2}\times35$ = 49.5 km $\rm s^{-1}$. We correct this 
and obtain the satellite velocity dispersion using
\begin{equation}
\sigma_{\rm s} = \sqrt{\sigma_{\rm v} ^{2} - \sigma_{\rm e} ^{2}}\,. 
\end{equation}

To compare with \cite{Zhang2022}, we use the same reference galaxy sample
to select satellite candidates.
This sample, selected from NYU-VAGC sample dr72\citep{Blanton-05a},
contains 510,605 galaxies, with $r$-band Petrosian apparent magnitude $r<$ 17.6, 
$r$-band Petrosian absolute magnitude in the range of $-24< M_{^{0.1}\rm r}<-16$, 
and the redshift in the range of $0.01<z<0.2$. 
Here, $M_{^{0.1}\rm r}$ is the $r$-band Petrosian absolute magnitude with $K+E$ 
corrected to the value at $z =0.1$ \citep[see][for more details]{WangL2019}. 

\subsection{The combined halo mass proxies}\label{sec_new_proxy}

In this section, we introduce our method for deriving new halo mass proxies
by linearly combining two proxies presented above. Since we have no 
$\sigma_{\rm s}$ measurements for individual galaxies, we only 
use $M_*$, $\sigma_*$ and $M_{\rm AM}$ to design new proxies. 
The three parameters have very different dynamical ranges, it
is thus necessary to scale them before the combination. 
For a proxy D, we design a new parameter,
\begin{equation}
    \rm Rank_D = \frac{D - min(D)}{max(D) - min(D)}\\,
\end{equation}
where $\rm min(D)$ and $\rm max(D)$ represent the minimum and 
maximum values of the proxy, respectively. A new proxy can then be designed 
by combining two proxies D and E as,
\begin{equation}\label{eq_proxy}
    \rm Pr(D, E, \emph{p}) = (1-\emph{p})*Rank_D + \emph{p}*Rank_E\\, 
\end{equation}
where $p$ is in the range of $0 \le p \le 1$. With $p=0$, $\rm Pr(D, E, \emph{p})$ 
is equivalent to the proxy D, while with $p=1$, $\rm Pr(D, E, \emph{p})$ is 
equivalent to the proxy E. Since $M_{\rm AM}$ is used, we 
use the B-series samples to examine the new proxies.

One way to evaluate the performance of a proxy is to check the dependence of 
the $M_{\rm h}$-proxy scaling relation on other galaxy and group properties, 
such as $N_{\rm sat}$ and SFR. We choose to use the differences in the 
$M_{\rm h}$-proxy scaling relations between BC0 and BC3 subsamples and 
between BSF and BQ subsamples to quantify the dependencies on 
$N_{\rm sat}$ and SFR, respectively. 
We first divide the BTotal sample into $N_{\rm b}=8$ equally-sized bins 
according to a new proxy, $ \rm Pr(D, E, \emph{p})$. Then we select galaxies 
that belong to BC0, BC3, BSF and BQ subsamples and derive
their halo masses at each $\rm Pr(D, E, \emph{p})$ bin based on weak lensing, respectively. 
We now have $M_{\rm h}$-$ \rm Pr(D, E, \emph{p})$ relations for BC0, BC3, BSF 
and BQ subsamples. We use the following formula to calculate the difference 
between two scaling relations,
\begin{equation}\label{eq_d_parameter}
    d_{\rm para} = \frac{1}{N_{\rm b}}\Sigma_{i=1}^{N_{\rm b}}\,\frac{(\log M_{i,1}-\log M_{i,2})^2}{err_{i,1}^2 + err_{i,2}^2}\\,
\end{equation}
where $\log M_{i,1}$, $err_{i,1}$, $\log M_{i,2}$ and $err_{i,2}$ are the halo masses 
and their uncertainties at the $i$th bin for the two scaling relations, respectively. 
Sometimes, the mean $ \rm Pr(D, E, \emph{p})$ at the largest or smallest 
$ \rm Pr(D, E, \emph{p})$ bin is very different between BC0 and BC3 
(or between BSF and BQ). In order to compare the two scaling relations in a fair way, 
the halo mass and its uncertainty used in Equation \ref{eq_d_parameter} for 
BC3 and BSF are calculated using the linear interpolation method. 
To evaluate the uncertainty of $d_{\rm para}$, we use the Monte Carlo method 
to generate $\log M_{i,1}$ and $\log M_{i,2}$ for each $ \rm Pr(D, E, \emph{p})$ bin, 
which follow Gaussian distributions with a mean equal to the weak lensing mass 
and a dispersion equal to the uncertainty. These new parameters can be used to 
calculate a new $d_{\rm para}$. This procedure is repeated 500 times and we 
take the dispersion of the 500 $d_{\rm para}$ as the uncertainty.
In the following, we use $d_{\rm SFR}$ and $d_{N_{\rm sat}}$ to represent the
dependencies of the scaling relation on SFR and $N_{\rm sat}$, respectively 
(Section \ref{sec_combined_proxy}).

\section{Halo mass - proxy scaling relations}\label{sec_scaling_relations}


\subsection{Stellar mass-halo mass scaling relation}\label{sec_SHMR}

\begin{figure*}
    \centering
    \includegraphics[scale=0.11]{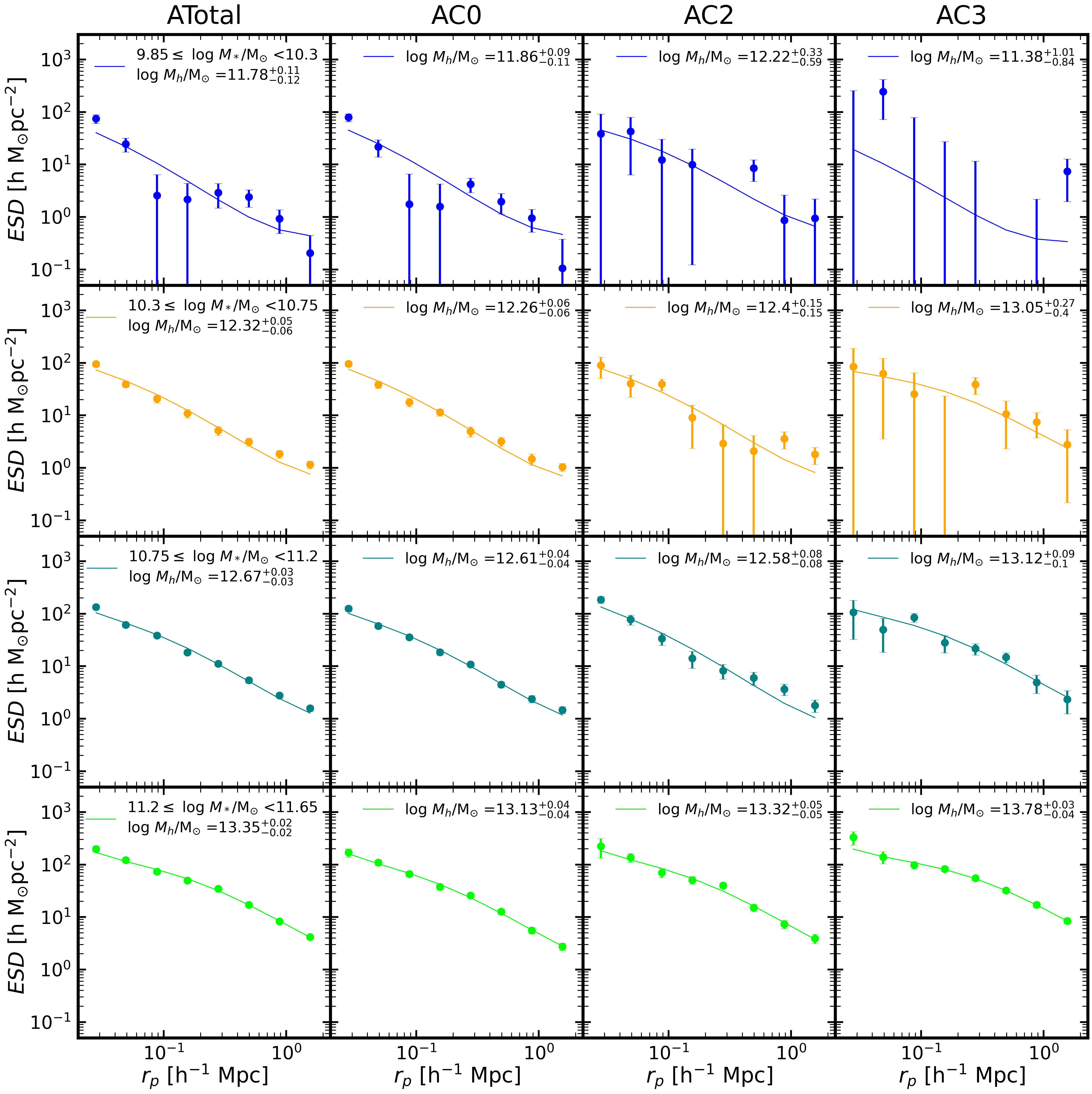}
    \caption{Excess surface density (ESD) profiles and the best fitting results. Different rows correspond to different $M_*$ bins, and different columns correspond to different galaxy samples, including ATotal, AC0, AC1, AC2 and AC3. The error bars correspond to the standard deviation of 150 bootstrap samples. The best-fitting $M_{\rm h}$ are also presented.}
    \label{fig_lensing}
\end{figure*}

\begin{figure*}
    \centering
    \includegraphics[scale=0.11]{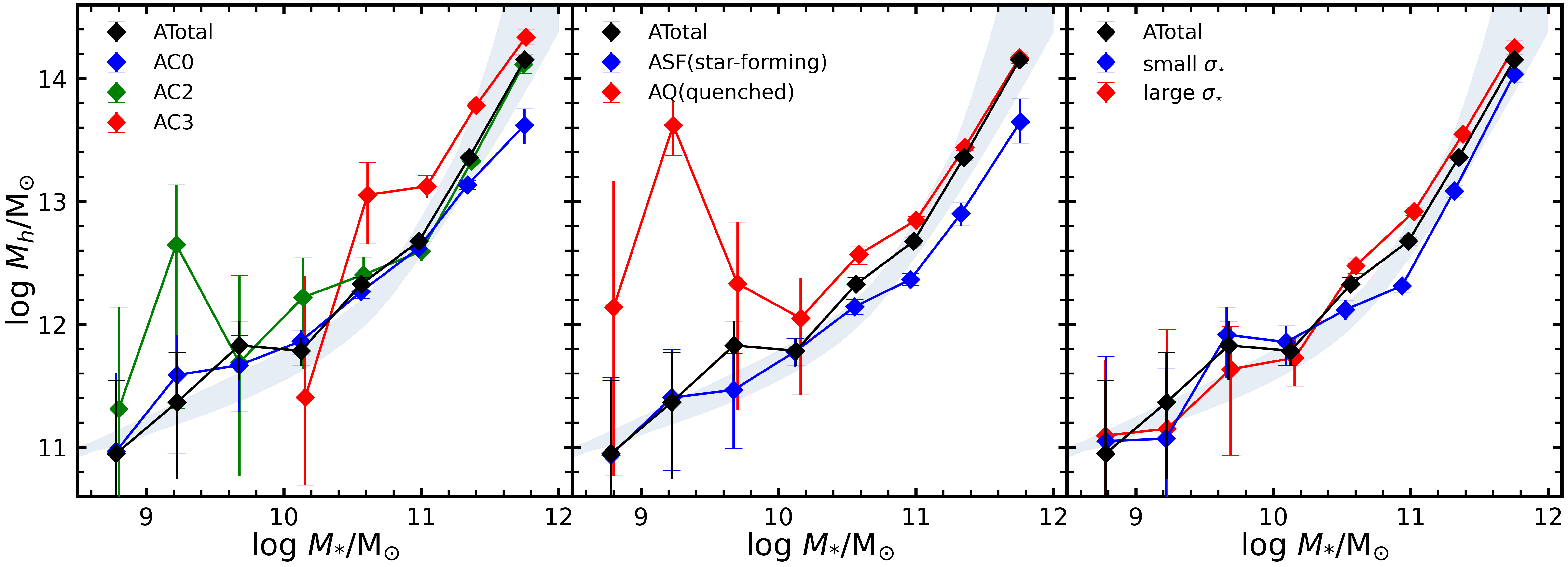}
    \caption{The dependence of SHMR on $N_{\rm sat}$(left), SFR(middle) and $\sigma_*$(right). The black lines in the three panels show the SHMR for ATotal sample. 
    The error bars reflect the 16\% and 84\% of the posterior distribution. The shaded region corresponds to the region covered by the SHMR curves published in \cite{Yang2009}, \cite{Moster2010}, \cite{Leauthaud2012}, \cite{Kravtsov2018}, and \cite{Behroozi2019}. }
    \label{fig_SHMR}
\end{figure*}

We divide galaxies from the ATotal sample with $M_* \ge 10^{8.5} \Msun$ into 8 $M_*$ bins, 
equally spaced in the logarithm of $M_*$ with a bin size of 0.45 dex. Then, we use 
methods presented in Section \ref{sec_measurements} to obtain the ESD and derive 
the halo mass for each $M_*$ bin. The ESD profiles and their best-fittings of 
four selected $M_*$ bins are presented in Figure \ref{fig_lensing}.
The resulting SHMR is shown in Figure \ref{fig_SHMR} as black solid diamonds. 
For comparison, we also show the SHMRs from the literature obtained using various methods, 
including galaxy groups, abundance matching, conditional luminosity function, 
weak lensing, and empirical models 
\citep[e.g.][]{Yang2009, Moster2010, Leauthaud2012, Kravtsov2018, Behroozi2019}. 
The SHMR for ATotal is well consistent with these results.

Many parameters are correlated with $M_{\rm h}$ and $M_*$. These correlations 
may reflect possible dependencies of the SHMR on these parameters. 
We first consider $N_{\rm sat}$. As shown in Section \ref{sec_galaxy_samples}, 
the ATotal sample can be split into AC0, AC2 and AC3 subsamples according 
to $N_{\rm sat}$. The ESDs and their best fitting profiles for these subsamples 
in four stellar mass bins are presented in Figure \ref{fig_lensing}. 
At a given $M_*$, the amplitude of the ESD signal increases with increasing group richness.
A direct comparison of the SHMRs can be found in the left panel of 
Figure \ref{fig_SHMR}, where only mass bins with more than 50 galaxies 
are presented. At the high mass end, the AC0 galaxies have a SHMR that tends 
to be slightly lower than that of ATotal, and AC3 galaxies tend to have the 
highest SHMR. We do not show the 
result for AC1 sample because AC1 is the sum of AC2 and AC3.

At given $M_*$, the halo mass is positively 
correlated with group richness \citep{Johnston2007, Andreon2008, Viola2015MNRAS.452.3529V} and the difference in $M_{\rm h}$ among samples is quite large. 
For the four most massive $M_*$ bins, the differences in $M_{\rm h}$ between 
AC0 and AC3 are in the range of 0.5-0.8 dex. 
The SDSS galaxy sample is a magnitude limited sample. Some groups in AC0 sample 
may have large intrinsic $N_{\rm sat}$, but their satellites are not detected 
due to the Malmquist bias. Such an effect dilutes the dependence on group richness. 
Therefore, the intrinsic dependence should be stronger than what we find here.

We then study the dependence of SHMR on SFR. The SHMR for star-forming (ASF) and quenched (AQ) galaxies are shown in the middle panel of Figure \ref{fig_SHMR}.
Overall, star-forming galaxies reside in halos with a mean halo mass much lower than quenched ones of the same $M_*$, consistent with previous studies \citep[e.g.][]{More2011, Rodriguez-Puebla2015ApJ...799..130R, Mandelbaum2016, Behroozi2019, Bilicki2021, Zhang2021, Zhang2022}. 
The quenched galaxies in the second lowest mass bin have a much higher halo mass compared to the star-forming ones. 
One possible reason is that a significant fraction of galaxies in this mass bin are satellites, which are misidentified as centrals by group finder. Because satellites usually reside in more massive halos than centrals of the same mass, the misidentification can lead to overestimation of halo mass. We examined their ESD profile and found a bump-like structure from 0.1-1 Mpc/h, which is similar to the lensing signal produced by satellite galaxies\citep{Yang2006, Lir2014}. Another possibility is that some of these galaxies reside in the splashback halos \citep{Wangk2023b}. Since splashback halos have ever entered massive halos, they are close to massive halos and thus have similar bump on their ESD. Moreover, these galaxies are subject to a variety of environmental effects and thus very likely to be quenched.

To investigate the dependence of SHMR on $\sigma_*$, we divide the ATotal sample in each 
stellar mass bin into small and large $\sigma_*$ subsamples by the median $\sigma_*$ of each 
$M_*$ bin. Their SHMRs are shown in the right panel of Figure \ref{fig_SHMR}. 
We can see that the SHMR has the strongest dependence on $\sigma_*$ in the middle stellar mass range. This dependence weakens towards the lower and higher stellar mass end.
We will come back to the correlations among $M_{\rm h}$, $M_*$ and $\sigma_*$ in the next subsection.


\subsection{Halo mass-galaxy central stellar velocity dispersion  scaling relation} \label{sec_Mh_g_sigma}

\begin{figure*}
    \centering
    \includegraphics[scale=0.11]{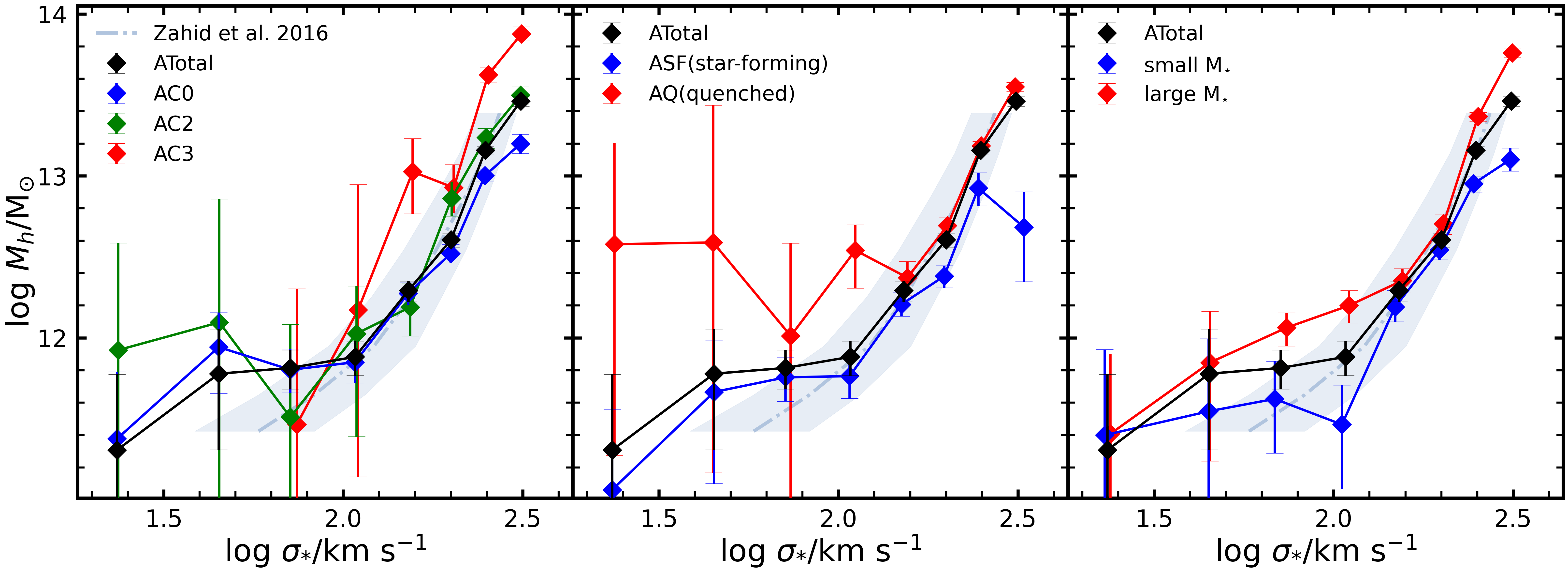}
    \caption{The dependence of the $M_{\rm h}$-$\sigma_*$ relation on $N_{\rm sat}$(left), SFR(middle) and $M_*$(right). The black lines in the three panels show the $M_{\rm h}$-$\sigma_*$ relation for ATotal sample.
  The light-blue dot-dashed line and the shaded region show the relation from \cite{Zahid2016} and its scatter. }
    \label{fig_hm_vds}
\end{figure*}

\begin{figure*}
    \centering
    \includegraphics[scale=0.13]{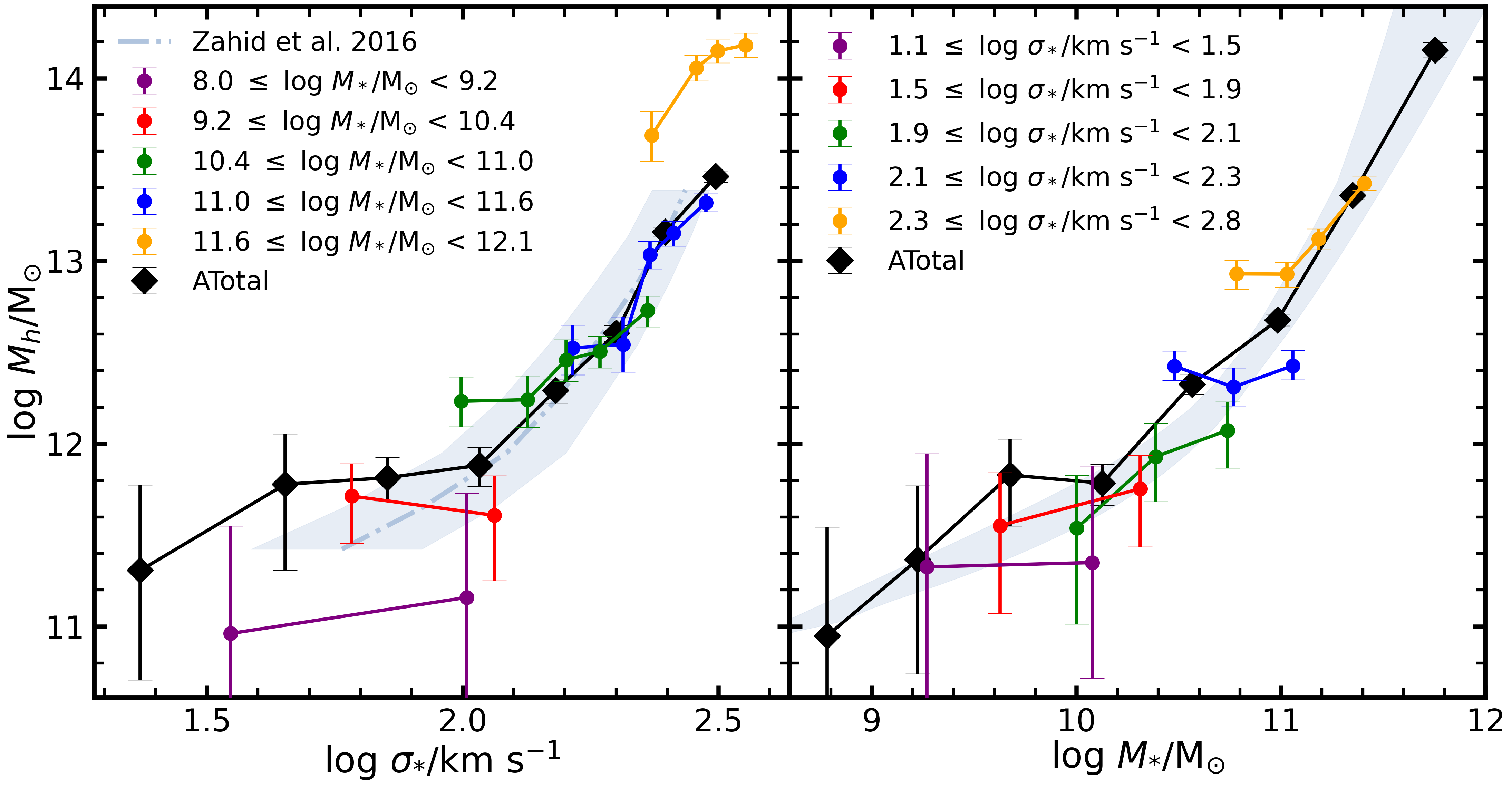}
    \caption{The dependence of $M_{\rm h}$-$\sigma_*$ relation on $M_*$ (left panel) and $M_{\rm h}$-$M_*$ relation on $\sigma_*$ (right panel). The black lines in the two panels show the results for ATotal sample. The color-coded lines in the left (right) panel show the results for $M_*$-controlled ($\sigma_*$-controlled) subsamples. Please see Section \ref{sec_Mh_g_sigma} for details. The dot-dashed line and the shaded region
    in the left panel are the same as those in Figure \ref{fig_hm_vds}. And the shaded region in the right panel is the same as that in Figure \ref{fig_SHMR}.}
    \label{fig_sm_vds_match}
\end{figure*}

In this subsection, we use the galaxy stellar velocity dispersion ($\sigma_*$) as a halo mass proxy and investigate the $M_{\rm h}$-$\sigma_*$ scaling relation.
The ATotal sample is divided into 8 $\sigma_*$ bins. The bin size varies with $\sigma_*$ to ensure that each bin has reliable $M_{\rm h}$ measurement as much as possible. The $M_{\rm h}$-$\sigma_*$ relation for the ATotal sample is shown in Figure \ref{fig_hm_vds} as black solid diamonds. We can see that galaxies with larger $\sigma_*$ reside in more massive halos.
For comparison, we also present the result of \cite{Zahid2016} who used the SHMR to assign the halo mass to each galaxy. Overall, our result is in good agreement with theirs.

We then examine the dependence of the $M_{\rm h}$-$\sigma_*$ relation on various parameters and only present the $\sigma_*$ bins with galaxy number greater than 50. We first investigate the dependence on $N_{\rm sat}$ (left panel of Figure \ref{fig_hm_vds}). Similar to the SHMR, the $M_{\rm h}-\sigma_*$ relation strongly depends on $N_{\rm sat}$. The uncertainties of halo masses at low $\sigma_*$ bins are very large, no clear $M_{\rm h}$-$N_{\rm sat}$ correlation can be found in this regime. However, at the four largest $\sigma_*$ bins where the halo masses are well constrained, $M_{\rm h}$ increases significantly with $N_{\rm sat}$ at fixed $\sigma_*$.  And the $M_{\rm h}$ difference between AC0 and AC3 is about 0.4-0.8 dex, similar to that for SHMR.

We now check the dependence of the $M_{\rm h}$-$\sigma_*$ relation on SFR(the middle panel of Figure \ref{fig_hm_vds}).
In the low $\sigma_*$ range ($\log\sigma_* < 2.1$), 
because of the small number of quenched galaxies, the uncertainties in the halo masses of quenched galaxies are very large.
In the middle $\sigma_*$ range ($2.1 <\log\sigma_* < 2.4$), the halo mass difference at a given $\sigma_*$ between star-forming and quenched galaxies is typically 0.2-0.3 dex, apparently smaller than the difference in the middle $M_*$ range, typically 0.4-0.5 dex.
This is not surprising, as $\sigma_*$ is the most important factor for star formation quenching\citep[see e.g.][]{Bluck2020b}. 
For the results in the largest $\sigma_*$ bin, the halo mass of the star-forming galaxies is significantly lower than that of the corresponding quenched galaxies. This result can be attributed to the fact that the star-forming galaxies contain a batch of galaxies with high star-forming efficiency\citep[see][]{Zhang2022}.
At a given $\sigma_*$, the halo masses of star-forming and quenched galaxies are still significantly different.
It suggests that other processes, likely related to halo mass, also play an important role in quenching star formation.

We then split each $\sigma_*$ bin into two equally-sized subsamples according to their $M_*$. The results for these subsamples are presented in the right panel of Figure \ref{fig_hm_vds}. 
The uncertainties in the two lowest $\sigma_*$ bins are too large to allow quantifying the difference. For larger $\sigma_*$, the dependence of the $M_{\rm h}$-$\sigma_*$ relation on $M_*$ is almost absent at $\log\sigma_*\sim2.3$ and becomes stronger towards lower and higher $\sigma_*$.
It is thus interesting to investigate the bivariate correlation in more detail.
We perform two tests. In the first test, we investigate the $M_{\rm h}$-$\sigma_*$ relation at a given $M_*$.  For galaxies in a given $M_*$ bin, we  split them 
into several(2 to 5, dependent on the weak lensing signal) equally-sized subsamples by $\sigma_*$. Our inspection shows that these subsamples with different $\sigma_*$ are significantly different in the $M_*$ distribution though they are selected within the same $M_*$ bin. To eliminate the potential bias, we re-select galaxies from these $\sigma_*$ subsamples so that they have similar $M_*$ distributions (see Appendix for the method). The re-selected subsamples are referred to as the $M_*$-controlled subsamples. In the second test, we investigate
the SHMR at a given $\sigma_*$. Similar to the first test, we
construct $\sigma_*$-controlled subsamples to eliminate the influence of $\sigma_*$. 

The left panel of Figure \ref{fig_sm_vds_match} shows the $M_{\rm h}$-$\sigma_*$ relations for $M_*$-controlled subsamples. 
At $\log\sigma_*<2.1$, $M_{\rm h}$ is almost independent of $\sigma_*$ but strongly dependent on
$M_*$. It suggests that for these galaxies, the observed $M_{\rm h}$-$\sigma_*$ relation is dominated by the $M_{\rm h}$-$M_*$ relation. At $2.1<\log\sigma_*<2.4$, we see a strong dependence of $M_{\rm h}$ on $\sigma_*$ even when $M_*$ is well controlled. Interestingly, 
galaxies in two different
$M_*$ bins(green and blue) follow almost the same trend, defined by the ATotal sample. It means that the SHMR in this range is driven by $\sigma_*$, totally opposite to the trend
seen in low-mass galaxies. 
At $\log\sigma_*>2.4$, we see the dependence on both $M_*$ and $\sigma_*$.

Similar conclusions can also be drawn from the SHMR results for the $\sigma_*$-controlled subsamples, as shown in the right panel of Figure \ref{fig_sm_vds_match}.
In the lowest three $\sigma_*$ bins, the SHMRs generally follow the same trend as that of the ATotal sample in the low stellar mass range ($\log M_*/\Msun < 10.4$). It means that $\sigma_*$ has a negligible impact on SHMR at the low $M_*$ end. In the middle stellar mass range ($10.4 < \log M_*/\Msun < 11.1$), the SHMRs in three $\sigma_*$ bins ($1.9<\log\sigma_*<2.8$) deviate from each other and all have very flat slopes. It means that $M_{\rm h}$ is independent of $M_*$ after controlling $\sigma_*$. Therefore, the SHMR in this mass range is mainly driven by the $M_{\rm h}$-$\sigma_*$ relation. In the massive stellar mass end ($\log M_*/\Msun > 11.1$), the SHMR in the largest $\sigma_*$ bin is almost consistent with the SHMR of ATotal sample. However, since there is only one $\sigma_*$ bin at this mass range, we can not disentangle the correlations with  $\sigma_*$ and $M_*$.
In general, these results are in good agreement with those from Figure \ref{fig_SHMR} and \ref{fig_hm_vds}.

\begin{figure}
    \centering
    \includegraphics[scale=0.14]{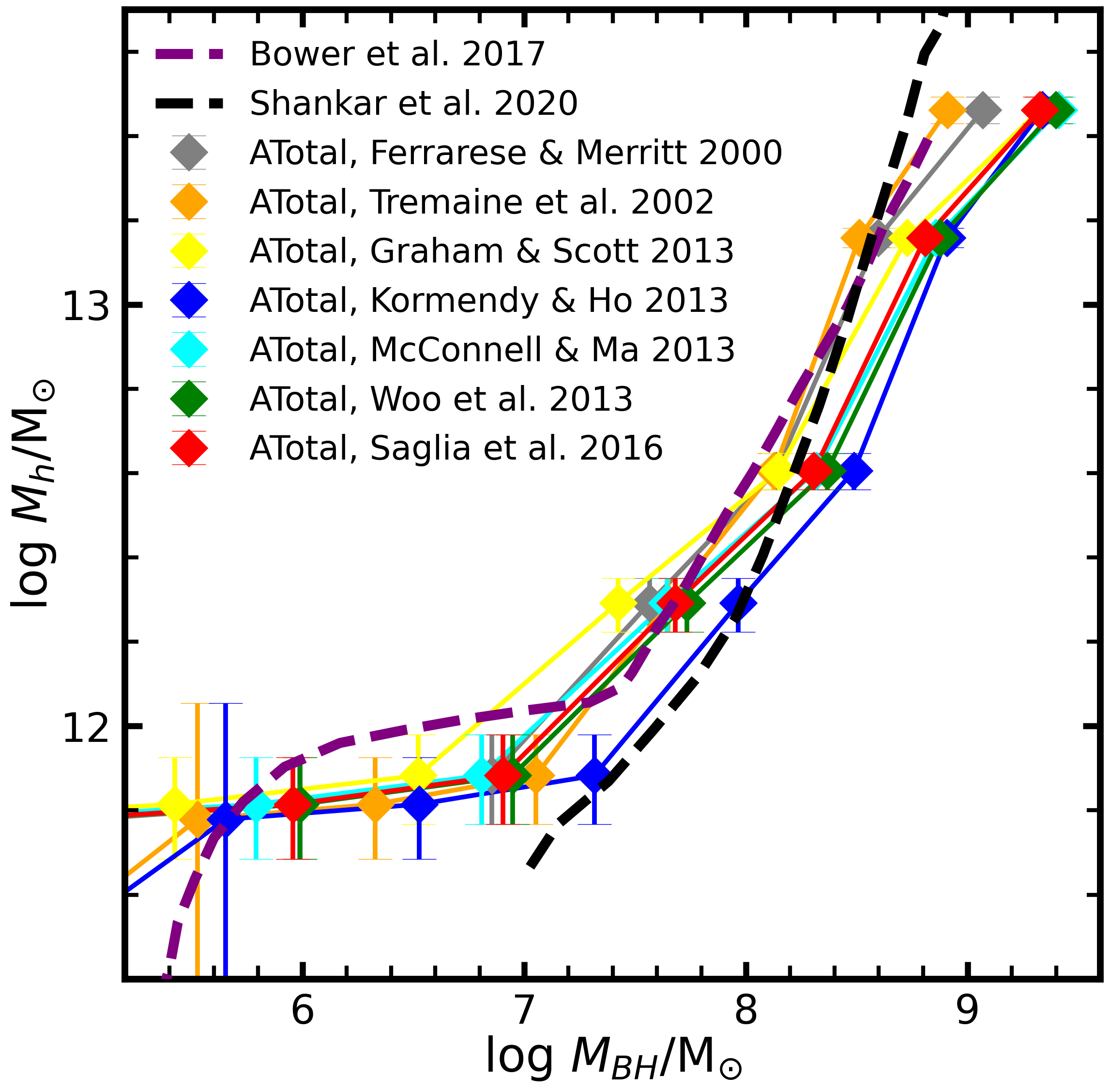}
    \caption{$M_{\rm h}$-$M_{\rm BH}$ relations. We apply seven $M_{\rm BH}$-$\sigma_*$ relations from the literature as indicated to convert the $M_{\rm h}$-$\sigma_*$ relation into the $M_{\rm h}$-$M_{\rm BH}$ relations.
    For comparison, we show the model result of \cite{Bower2017} in purple dashed line and the observational result of \cite{Shankar2020} in black dashed line.}
    \label{fig_hm_Mbh}
\end{figure}

One important application of the $M_{\rm h}$-$\sigma_*$ relation is that
it can be used to study the connection between dark matter halos and supper-massive black holes in galaxy center, which are thought to play an essential role in galaxy evolution.
It is well known that $\sigma_*$ is tightly
correlated with the black hole mass ($M_{\rm BH}$).
We can convert the $M_{\rm h}$-$\sigma_*$ relation into
$M_{\rm h}$-$M_{\rm BH}$ relation using the $M_{\rm BH}$-$\sigma_*$ relation. Here, we adopt seven $M_{\rm BH}$-$\sigma_*$ relations\citep[][]{Ferrarese2000, Tremaine2002, Graham2013, Kormendy2013, McConnell2013, Woo2013, Saglia2016}, which are derived from different galaxy samples. The resultant $M_{\rm h}$-$M_{\rm BH}$ relations are shown in Figure  \ref{fig_hm_Mbh}.

Though the uncertainty in the $M_{\rm BH}$-$\sigma_*$ relation
leads to a large uncertainty in the $M_{\rm BH}$ estimation,
all of these results follow a general trend.
When $M_{\rm BH}$ increases rapidly from $10^{5.0}$ to $10^{7.4}\Msun$,
$M_{\rm h}$ only increases about 0.2 dex, from $10^{11.8}$ to $10^{12.0}\Msun$.
It hints at a rapid growth of black holes as their host halos reach $\log M_{\rm h}/\Msun \sim 12.0$, consistent with that AGNs usually reside in halos of $\log M_{\rm h}/\Msun\sim12.0$\citep[e.g.][]{Zhang2021}.
At $\log M_{\rm h}/\Msun>12.0$, 
when $M_{\rm BH}$ increases from $10^{7.4}$ to $10^{9}\Msun$, $M_{\rm h}$ increases from $10^{12.0}$ to $10^{13.4}\Msun$, indicating a nearly linear correlation between $M_{\rm BH}$ and $M_{\rm h}$.  Our $M_{\rm h}$-$M_{\rm BH}$ relation shows a transition in black hole growth at 
 $\log M_{\rm h}/\Msun\sim12.0$ and $\log M_{\rm BH}/\Msun\sim7.4$. 

It is possible that the independence of
$M_{\rm h}$ on $M_{\rm BH}$ at low mass end is caused by the relatively large uncertainties of the low $\sigma_*$.
Because large uncertainties can erase the intrinsic difference among galaxies in several low $\sigma_*$ bins.
To check this possibility, we investigate the errors of $\sigma_*$ measurement, taken from NYU-VAGC, in the four lowest $\sigma_*$ bins of the ATotal sample. Our tests show that the four lowest $\sigma_*$ bins (at least the second to fourth bin) have significantly different intrinsic $\sigma_*$ considering the errors. It is also supported by the fact that the $M_*$ distributions in the four bins are significantly different.
Another important factor is the uncertainty of $M_{\rm BH}$-$\sigma_*$ relation at the low mass end. We adopt seven relations, which are obtained using different galaxy sample and have different slopes. 
Similar transition is seen in all results. It is expected because the transition originates from the $M_{\rm h}$-$\sigma_*$ relation where the transition is clearly presented. 
We also calculate the $M_{\rm BH}$-$M_{\rm h}$ relation for early-type galaxies and find a similar transition. These tests indicate the uncertainties in $\sigma_*$ measurement and $M_{\rm BH}$-$\sigma_*$ relation do not significantly change our conclusion.


For comparison, we show the $M_{\rm h}$-$M_{\rm BH}$ relation obtained by \cite{Shankar2020}. 
\cite{Shankar2020} estimated $M_{\rm h}$ using the abundance matching technique and $M_{\rm BH}$ using the $M_*$-$M_{\rm BH}$ relation for quiescent galaxies\citep{Saintonge2016}. 
In general, our predictions are consistent with theirs, but lower at high mass and higher at low mass.
\cite{Shankar2020} also provided a $M_{\rm h}$-$M_{\rm BH}$ relation based on
a so-called bias-corrected galaxy sample, which significantly deviates from ours.
Because we adopt different methods to derive $M_{\rm h}$ and $M_{\rm BH}$, it is difficult to determine the exact cause of this discrepancy. We also show the result of an analytic model developed by \cite{Bower2017}. 
Their model prediction is in good agreement with our observational result at $\log M_{\rm BH}/\Msun>6$. They suggested that when $M_{\rm h}$ is close to $10^{12}\Msun$, star-formation feedback can not effectively drive outflow and black holes start to grow rapidly. When $M_{\rm h}>10^{12}\Msun$, the ISM will eventually be consumed by star formation or expelled by AGN feedback, and the black hole growth rate is reduced. \cite{Bower2017} assumed that the further growth of black holes is limited by the binding energy within the cooling radius of their host halos, which leads to the linear relationship between the halo mass and the black hole mass at $\log M_{\rm h}/\Msun>12$. 
Apparently, the $M_{\rm h}$-$\sigma_*$ and $M_{\rm h}$-$M_{\rm BH}$ relations
contain valuable information about galaxy formation and evolution.
We will come back to this issue in the near future (Li. et al. in preparation).

\subsection{Halo mass-abundance-matching mass scaling relation}\label{sec_MAM}

\begin{figure*}
    \centering
    \includegraphics[scale=0.13]{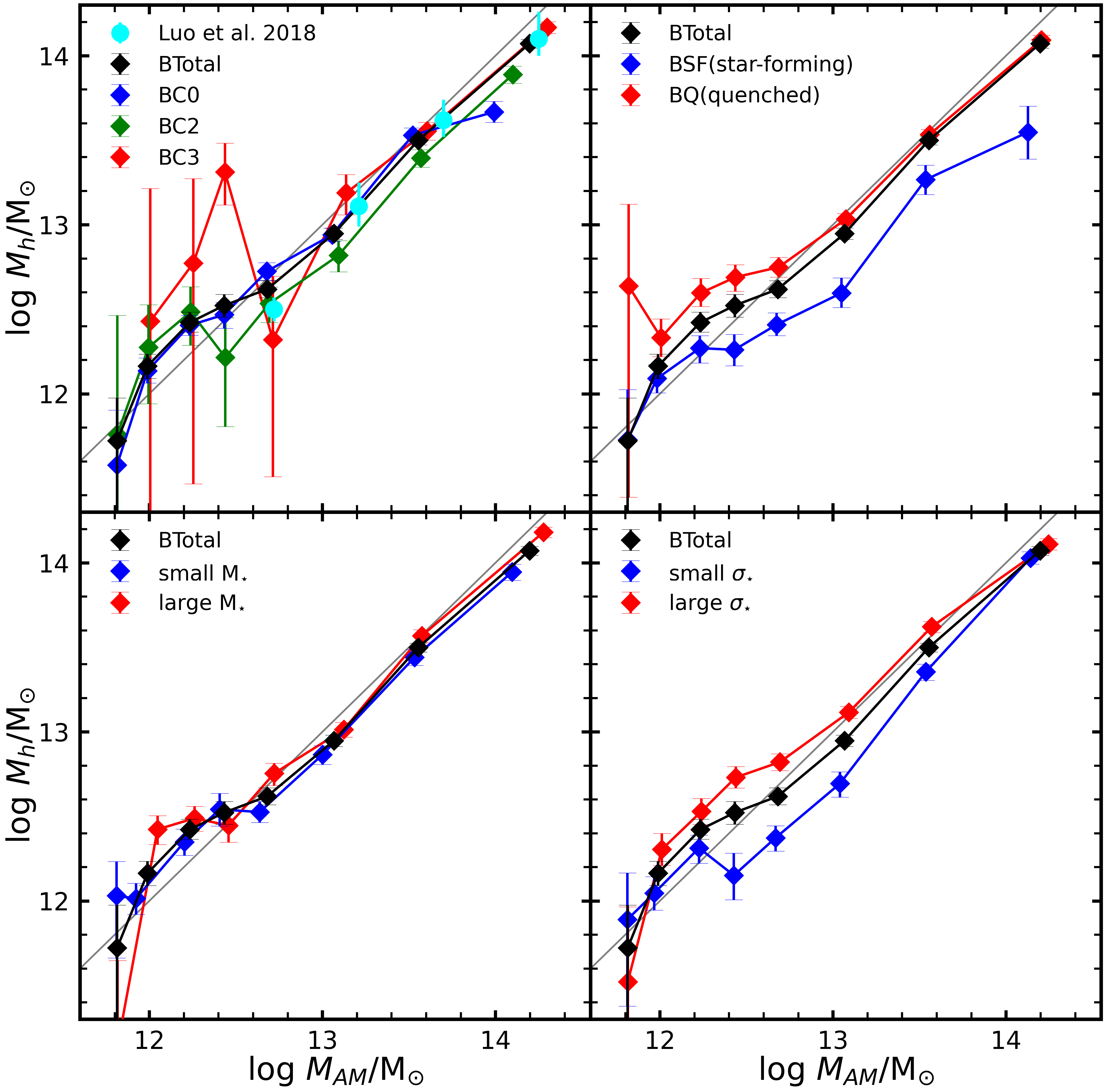}
    \caption{The dependence of the $M_{\rm h}$-$M_{\rm AM}$ relation on $N_{\rm sat}$ (upper-left), SFR(upper-right), $M_*$(lower-left) and $\sigma_*$(lower-right).
    The cyan circles show the results from \cite{Luo2018ApJ}, and the gray line in each panel shows the one-to-one relation.}
    \label{fig_hm_Mhs}
\end{figure*}

In this subsection, we adopt the abundance matching halo mass (hereafter $M_{\rm AM}$) based on group stellar mass as the halo mass proxy. The following tests are based on the B-series samples. We divide the BTotal sample into 8 $M_{\rm AM}$ bins, the sizes of which are varied to
obtain reliable halo mass measurements as far as possible. The resultant $M_{\rm h}$-$M_{\rm AM}$ relation is shown in Figure \ref{fig_hm_Mhs} as black solid diamonds. For comparison, we also present the $M_{\rm h}$-$M_{\rm AM}$ relation from \cite{Luo2018ApJ}, shown as the cyan circles. Our result is in excellent agreement with that of \cite{Luo2018ApJ}. The $M_{\rm h}$-$M_{\rm AM}$ relation is very close to a one-to-one relation. At $\log M_{\rm h}/\Msun>12.5$, the weak lensing measured halo mass is slightly smaller than the abundance matching mass. At $\log M_{\rm h}/\Msun<12.5$, $M_{\rm h}$ becomes slightly larger than $M_{\rm AM}$. 

We then examine the dependence of the $M_{\rm h}$-$M_{\rm AM}$ relation on various parameters. The upper-left panel shows the dependence on group richness. Different from the SHMR and $M_{\rm h}$-$\sigma_*$ relation, the dependence of the $M_{\rm h}$-$M_{\rm AM}$ relation on $N_{\rm sat}$ is much weaker. $M_{\rm AM}$ is estimated based on the sum of stellar masses of group member galaxies \citep[see][ for the details]{Yang2007}. Therefore, the satellite information has already been involved in the estimation of $M_{\rm AM}$.  At low mass, the result for the BTotal is
dominated by the BC0 galaxies, which have no satellite. It means
that the $M_{\rm h}$-$M_{\rm AM}$ relation is approximately equivalent to the $M_{\rm h}$-$M_*$ relation.
It might explain the small change in the $M_{\rm h}$-$M_{\rm AM}$ relation at $\log M_{\rm h}/\Msun\sim 12.5$. 
We also divide galaxies in each $M_{\rm AM}$ bin into two equally-sized subsamples according to their $M_*$. The $M_{\rm h}$-$M_{\rm AM}$ relations
for the small and large $M_*$ galaxies are presented in the lower-left panel of Figure \ref{fig_hm_Mhs}. As one can see, they both closely follow the overall $M_{\rm h}$-$M_{\rm AM}$ relation and exhibit almost no difference. It means that the $M_{\rm h}$-$M_*$
relation can be fully explained by the $M_{\rm h}$-$M_{\rm AM}$ relation. Therefore, $M_{\rm AM}$ is a more powerful proxy than $M_*$.
It is not surprising because $M_{\rm AM}$ is estimated based on the
group stellar mass, which also includes the contribution of central galaxies.

The $M_{\rm h}$-$M_{\rm AM}$ relations for star-forming and quenched galaxies are shown in the upper-right panel of Figure \ref{fig_hm_Mhs}. The halo masses for the quenched galaxies are apparently higher than the corresponding star-forming galaxies over all $M_{\rm AM}$ range in consideration. The halo mass differences are rather large, ranging from $\sim$ 0.3 to 0.9 dex, comparable to the difference for the $M_{\rm h}$-$M_*$ relation and larger than that for the $M_{\rm h}$-$\sigma_*$ relation. 
The dependence of the $M_{\rm h}$-$M_{\rm AM}$ relation on $\sigma_*$ is 
also significant (lower-right panel). 
At low $M_{\rm AM}$, the dependence on $\sigma_*$ is weak. At middle $M_{\rm AM}$, the dependence becomes very pronounced, with large $\sigma_*$ galaxies residing in more massive halos than the corresponding small $\sigma_*$ galaxies. At high $M_{\rm AM}$, this dependence becomes weak, and almost disappears in the largest $M_{\rm AM}$ bin. 
Overall, $M_{\rm AM}$ is a better proxy than $M_*$ and $\sigma_*$. However, the dependencies on SFR and $\sigma_*$ suggest that the $M_{\rm h}$-$M_{\rm AM}$ scaling relation is still sensitive to galaxy formation processes. 

\subsection{Halo mass-satellite velocity dispersion scaling relation} \label{sec_Mh_sigmas}

\begin{figure*}
    \centering
    \includegraphics[scale=0.24]{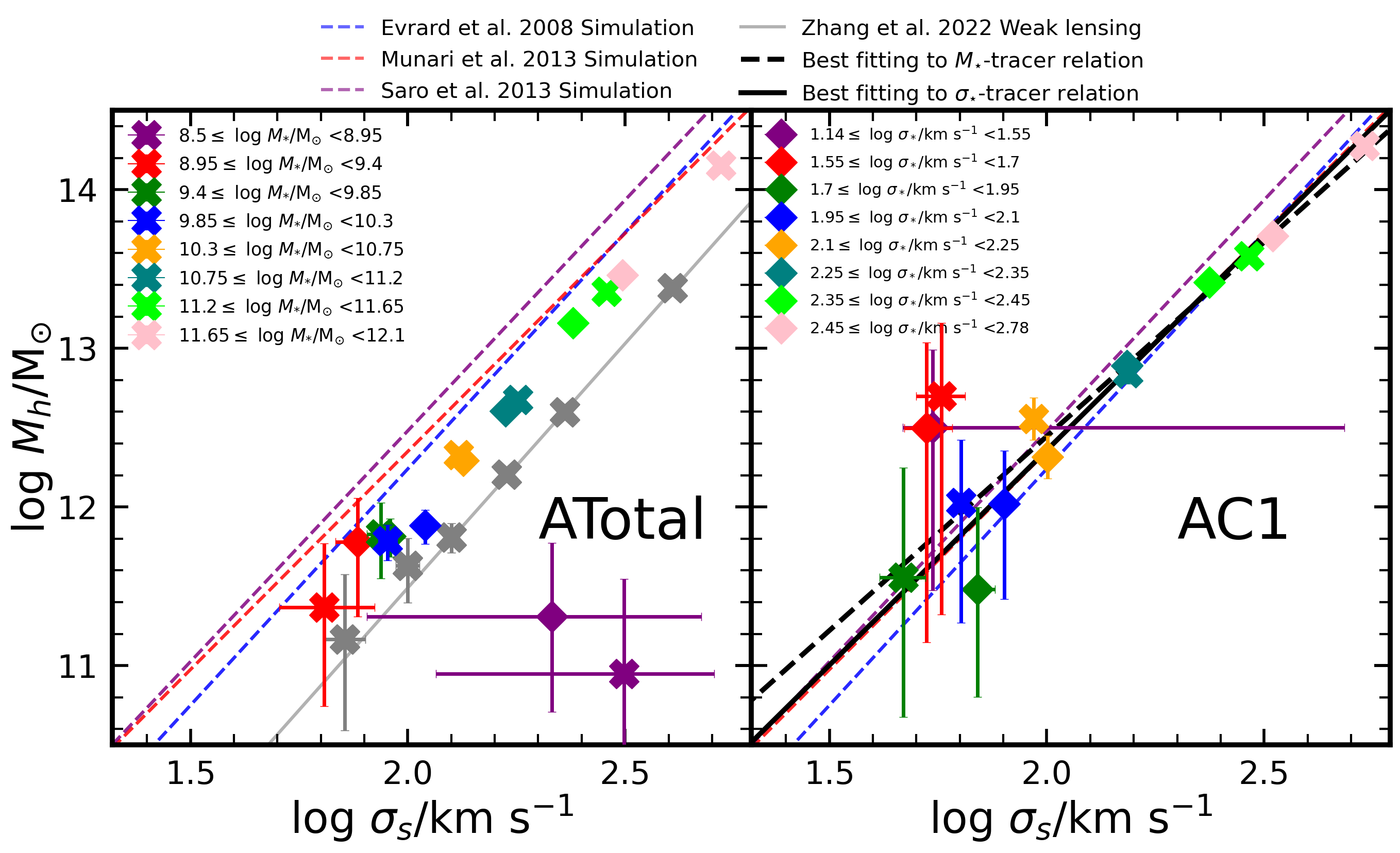}
    \caption{The $M_{\rm h}$-$\sigma_{\rm s}$ relations 
    using $M_*$ (crosses) or $\sigma_*$ (diamonds) as a tracer.
    Different colors represent different $M_*$ or $\sigma_*$ bins
    as labeled in the two panels. The left and right panels show
    the results based on Atotal and AC1 samples, respectively.  The black dashed and solid lines are the best-fittings to $M_*$- and $\sigma_*$-tracer relations, respectively. 
    The red, purple and blue dashed lines show the simulation results. The gray solid crosses are taken from \cite{Zhang2022} and the gray line is its best-fitting (left panel). Please see Section \ref{sec_Mh_sigmas} for the details.}
    \label{fig_smh_obs}
\end{figure*}

The halo mass-satellite velocity dispersion ($\sigma_{\rm s}$) scaling relation has been
widely studied in the literature \citep[e.g.][]{Rines2006, Yang2007, Hoekstra2007, Evrard2008, Munari2013, Rines2013, Saro2013, Ruel2014, Gonzalez2015, Han2015, Viola2015MNRAS.452.3529V, Rines2016, Abdullah2020, Gonzalez2021, Rana2022}. Most of these studies used the $\sigma_{\rm s}$ data measured for individual galaxy groups/clusters even though these groups have only a few member galaxies. 
We, however, adopt a stacking method to measure the average $\sigma_{\rm s}$ for a given galaxy sample (see the method description in Section \ref{sec_sk}). Our method needs a halo mass tracer to
divide a galaxy sample into several bins. As shown in the above sections, $M_*$, $\sigma_*$ and $M_{\rm AM}$ are all strongly correlated with halo mass. We thus decide to use these
three quantities as halo mass tracers. When we use $M_*$ and $\sigma_*$ as tracers, the
tests are performed on the A-series samples. When $M_{\rm AM}$ is considered, the tests are performed on the B-series samples. We divide the galaxy samples into 8 bins according to these tracers. 
Since both the weak lensing and satellite kinematics techniques are performed using the stacking method, the uncertainties are very large when the sample size is small. For clarity, we only present the results with galaxy number greater than 100.

\begin{table*}
\caption{The fitting parameters for $M_{\rm h}$-$\sigma_{\rm s}$ relation}
\label{tab_Mh_sigmas}
\centering
\begin{threeparttable}
\setlength{\tabcolsep}{5.5mm}{
\begin{tabular}{c c c c c}
\hline  
Reference & $b$\tnote{(a)} & $k$\tnote{(b)} & Tracer\tnote{(c)} & Data\tnote{(d)}\\
\hline  
\cite{Evrard2008}       &  6.28  &  2.98 & -- & simulation\\
\cite{Munari2013}       &  6.85  &  2.75 & -- & simulation\\
\cite{Saro2013}         &  6.66  &  2.91 & -- & simulation\\
\cite{Han2015}          &  7.77$^{+0.08}_{-0.08}$  &  2.17$^{+0.34}_{-0.34}$ & 
$\sigma_{\rm s}$ & observation\\
\cite{Viola2015MNRAS.452.3529V}  &  9.07$^{+0.06}_{-0.07}$  &  1.89$^{+0.27}_{-0.27}$ & 
$\sigma_{\rm s}$ & observation\\
\cite{Rana2022}         &  10.03$^{+0.02}_{-0.02}$  &  1.52 $^{+0.1}_{-0.1}$ & 
$\sigma_{\rm s}$ & observation\\
\cite{Zhang2022}        &  5.33$^{+0.11}_{-0.11}$  &  3.08$^{+0.05}_{-0.05}$ & 
$M_{\rm *}$ & observation\\
\hline
Sample & $b$\tnote{(a)} & $k$\tnote{(b)} & Tracer\tnote{(c)} & Data\tnote{(d)}\\
\hline
AC1                  & 7.55$^{+0.06}_{-0.06}$ & 2.45$^{+0.03}_{-0.03}$ & $M_*$ & observation\\
AC1                  & 6.94$^{+0.09}_{-0.09}$ & 2.71$^{+0.04}_{-0.04}$ & $\sigma_*$ & observation\\
BC1                  & 7.31$^{+0.06}_{-0.06}$ & 2.57$^{+0.02}_{-0.02}$ & $M_{\rm AM}$ & observation\\
BC1, star-forming    & 6.12$^{+0.2}_{-0.22}$ & 3.08$^{+0.1}_{-0.09}$ & $M_{\rm AM}$ & observation\\
BC1, quenched        & 7.23$^{+0.07}_{-0.07}$ & 2.61$^{+0.03}_{-0.03}$ & $M_{\rm AM}$ & observation\\
BC2                  & 6.38$^{+0.13}_{-0.14}$ & 3.02$^{+0.06}_{-0.06}$ & $M_{\rm AM}$ & observation\\
BC3                  & 6.95$^{+0.12}_{-0.12}$ & 2.72$^{+0.05}_{-0.05}$ & $M_{\rm AM}$ & observation\\
\hline
\end{tabular}}

\begin{tablenotes}
\footnotesize
\item[(a)] Zero-point($b$) in Equation \ref{eq_fitting}.
\item[(b)] Slope($k$) in Equation \ref{eq_fitting}.
\item[(c)] Samples are divided into subsamples according to the tracer for fitting.
\item[(d)] The origin of the data used for fitting.
\end{tablenotes}
\end{threeparttable}
\end{table*}

We first show the results using $M_*$ and $\sigma_*$ as tracers in Figure \ref{fig_smh_obs}. They are referred to as the $M_*$-tracer (solid crosses) and $\sigma_*$-tracer (solid diamonds) relations, respectively. 
For reference, we also present the scaling relations obtained from simulations\citep[][]{Evrard2008, Munari2013, Saro2013}. 
The blue dashed line shows the relation from \cite{Evrard2008} that used dark matter particles to calculate the satellite velocity dispersion. They obtained a slope of 2.98, in good agreement with the virial scaling relation(a slope of 3). 
The red dashed line shows the result from a hydro-dynamical simulation with star formation and AGN feedback\citep{Munari2013}, in which the simulated galaxies are used to calculate the dispersion. They got a slope of 2.75, slightly less than 3. 
The purple dashed line shows the relation from \cite{Saro2013}, which has a slope of 2.91. They used
galaxies yielded by a semi-analytic galaxy formation model
and only considered galaxy clusters. 
The best-fitting parameters for the three results are listed in Table \ref{tab_Mh_sigmas}. Note that their halo masses have already been converted into our halo mass definition.

The results from \cite{Zhang2022} are also presented in gray color, which used $M_*$ as a tracer and adopted a larger $\Delta v$ cut of $3v_{\rm 200m}$. They obtained a slope close to the simulations and an amplitude, however, much lower than the simulations (Table \ref{tab_Mh_sigmas}). 
Although $\sigma_{\rm s}$ in \cite{Zhang2022} is overestimated, the halo mass estimated using the $M_{\rm h}$-$\sigma_{\rm s}$ relation is still reliable.
Because the $M_{\rm h}$-$\sigma_{\rm s}$ relation is calibrated by using weak lensing measured halo mass. The reasons for the large discrepancy between the $M_{\rm h}$-$\sigma_{\rm s}$ relation of \cite{Zhang2022} and the simulation relations are twofold: (1) the contamination of interlopers as discussed in Section \ref{sec_sk}; (2) the weighting bias discussed in \cite{vandenBosch2004}.
To check the first issue, we adopt a $\Delta v$ cut of $1.5v_{\rm 200m}$ and present the results of ATotal sample in the left panel of Figure \ref{fig_smh_obs}. As one can see, a small $\Delta v$ cut can significantly reduce the contamination.
However, the discrepancy is still severe. AC0 galaxies have no satellite and thus contribute little to the $\sigma_{\rm s}$ measurement, but they have on average lower $M_{\rm h}$ than AC1 galaxies and thus significantly lower the mean halo mass. It means that AC0 galaxies have different weighting factors in the measurements of $M_{\rm h}$ and $\sigma_{\rm s}$.
To reduce the weighting bias, we exclude AC0 galaxies and apply our method to AC1 galaxies only. The new scaling relations lie very close to the simulation results, as shown in the right panel of Figure \ref{fig_smh_obs}.

\begin{figure*}
    \centering
    \includegraphics[scale=0.20]{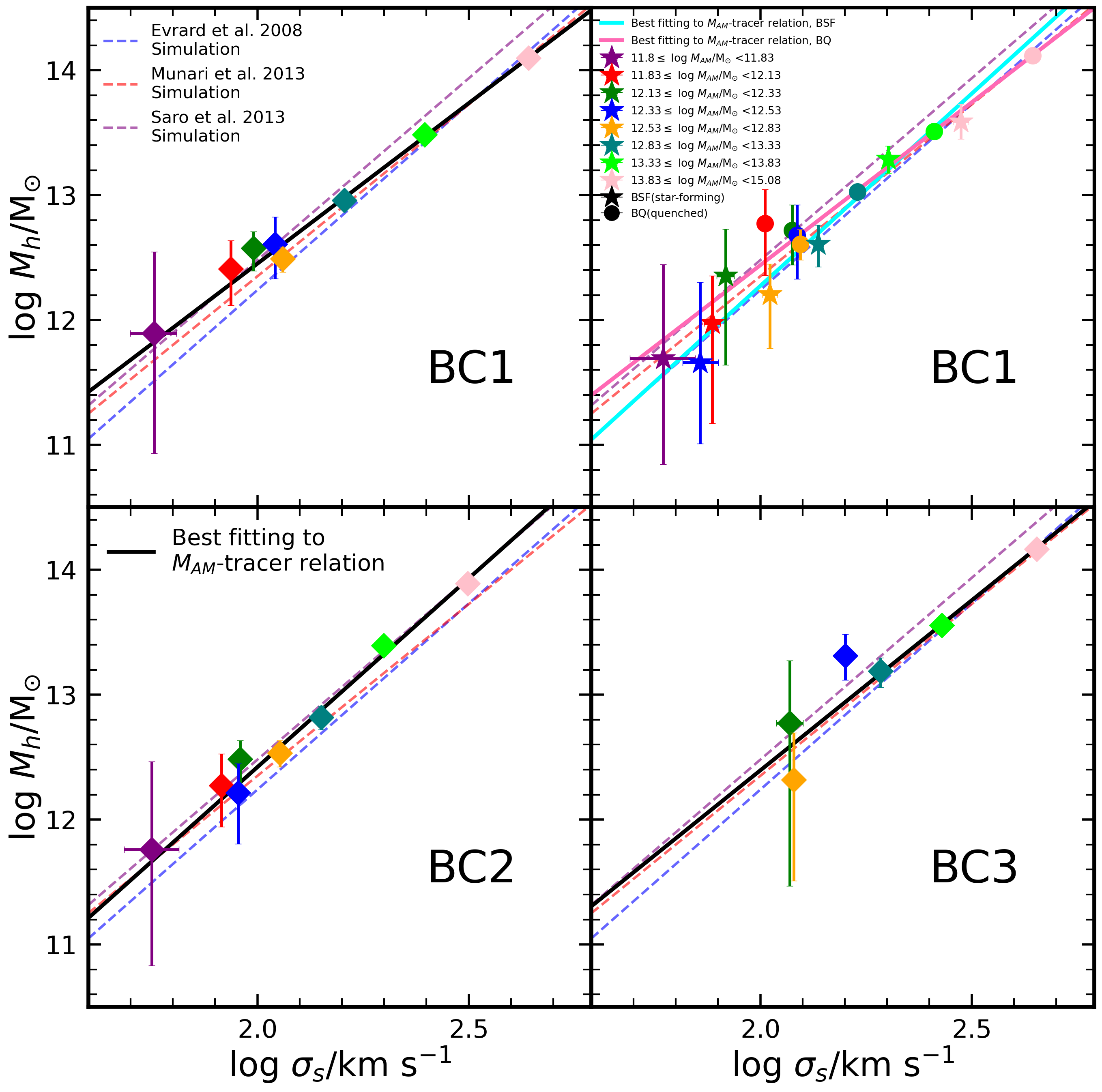}
    \caption{The $M_{\rm h}$-$\sigma_{\rm s}$ relations using $M_{\rm AM}$ as a tracer. Different colors represent different $M_{\rm AM}$ bins as labeled in the upper-right panel. The upper-left, lower-left and lower-right panels show the results of BC1, BC2 and BC3, respectively. The upper-right panel shows the results for star-forming(solid stars) and quenched(solid circles) galaxies in BC1 sample. In each panel, the solid line is the best-fitting to the corresponding sample, and the red, purple and blue dashed lines are the relations from simulations.}
    \label{fig_smh_Mhs}
\end{figure*}

We use a power-law function to fit the data,
\begin{equation}
    \log M_{\rm h}/\Msun = k\log \sigma_{\rm s}/\kms +b.\\\label{eq_fitting}
\end{equation}
The best-fitting slopes are 2.45 and 2.71 for the $M_*$-tracer and $\sigma_*$-tracer relations, respectively (see Table \ref{tab_Mh_sigmas}). The best-fitting results are also presented in black lines in Figure \ref{fig_smh_obs}. We can see that the $\sigma_*$-tracer relation is perfectly consistent with the relation measured from hydrodynamical simulation\citep{Munari2013}. 
We also use $M_{\rm AM}$ as a tracer to divide BC1 sample into 8 $M_{\rm AM}$ bins. 
The obtained $M_{\rm h}$-$\sigma_{\rm s}$ scaling relation and its 
fitting result are presented in Figure \ref{fig_smh_Mhs} and Table \ref{tab_Mh_sigmas}. 
The $M_{\rm AM}$-tracer relation is also very similar to the simulation results.
It is almost the same as the results of \cite{Evrard2008} and \cite{Munari2013} at the massive end and slightly higher than them
at the low-mass end. The slope of the relation is 2.57 (Table \ref{tab_Mh_sigmas}), between those of the $M_*$- and $\sigma_*$-tracer relations. Our measurements are more similar
to the result of hydrodynamical simulation.
It might be a signal for velocity bias.

We then check the dependence of the $M_{\rm h}$-$\sigma_{\rm s}$ relation on other galaxy and group properties.  Here, we only present the $M_{\rm AM}$-tracer results. The results for BC2 and BC3 samples are shown in Figure \ref{fig_smh_Mhs} and Table \ref{tab_Mh_sigmas}. The relations for the two populations are both
similar to the simulation results. The relation for BC2(close to the simulation result of \cite{Saro2013}) is slightly higher than
that for BC3 (close to the result of \cite{Munari2013}) at high $\sigma_{\rm s}$. We also split the BC1 sample into star-forming
and quenched galaxies. The scaling relations for the two types of galaxies are also presented in Figure \ref{fig_smh_Mhs} and Table \ref{tab_Mh_sigmas}. In general,
the two relations are very similar and both are close to the simulation results. And the slope
for star-forming galaxies seems slightly steeper than that for quenched galaxies. Given the large uncertainties
for star-forming galaxies at low mass, the difference is not significant. 

\begin{figure}
\includegraphics[scale=0.23]{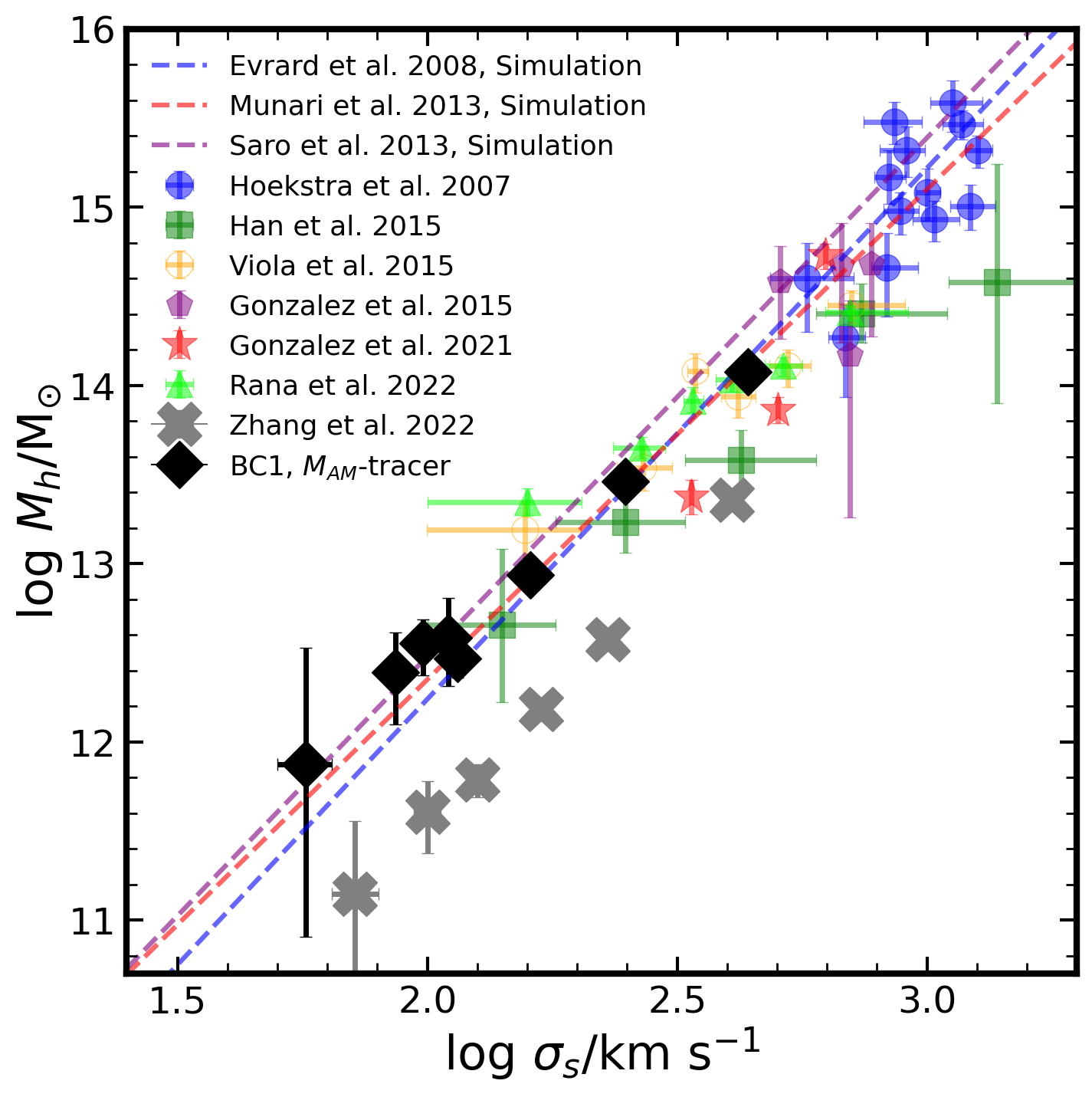}
\centering
\caption{The $M_{\rm h}$-$\sigma_{\rm s}$ relations from simulations and observations. As indicated in the figure, the color-coded dashed lines are the simulation results and the color-coded and gray symbols are the results from the literature with weak lensing measured $M_{\rm h}$. 
For comparison, we also show our results for BC1 sample with $M_{\rm AM}$ as the halo mass tracer in black solid diamonds.}
\label{fig_Mh_sigmas_compare}
\end{figure}

In addition, we perform a test with a double-Gaussian model\citep[see e.g.][]{More2011} to repeat Figure \ref{fig_smh_obs} and Figure \ref{fig_smh_Mhs}. In general, the obtained $M_{\rm h}$-$\sigma_{\rm s}$ relations from the double-Gaussian model are in agreement with the simulation relations too. However, they are more noisy compared to the results shown in this paper. It is likely that one of the two Gaussian components sometimes fit the interlopers rather than the satellites, resulting in a larger error bar in $\sigma_{\rm s}$. Therefore, we don't use the double-Gaussian model to analyze the satellite kinematics in this paper.

Finally, we compare our $M_{\rm AM}$-tracer $M_{\rm h}$-$\sigma_{\rm s}$ relations of BC1 sample with previous observational studies based on weak lensing in Figure \ref{fig_Mh_sigmas_compare}. \cite{Hoekstra2007} and \cite{Gonzalez2015} focused on
individual galaxy clusters. In general, these clusters lie around
the simulation relations with relatively large scatter. 
\cite{Han2015}, \cite{Viola2015MNRAS.452.3529V}, \cite{Gonzalez2021}, and \cite{Rana2022} calculated $\sigma_{\rm s}$ for individual
groups with more than 3 to 5 galaxy members and then divided their group sample into subsamples according to $\sigma_{\rm s}$. Therefore, they all used $\sigma_{\rm s}$ as a tracer.
We list their best fitting parameters (if available) in Table \ref{tab_Mh_sigmas}. These researches usually got slopes ($\sim 2$ or $<2$) that are much flatter than the simulation slopes.
Some of them got amplitudes lower than the simulations. 
Various possible interpretations for the discrepancies have been proposed, such as
rich group selection bias, interlopers, velocity bias and non-equilibrium systems.

Different from these studies, we divide galaxies by their $M_*$, $\sigma_*$ or $M_{\rm AM}$ rather than $\sigma_{\rm s}$. 
After carefully dealing with interlopers and weighting bias, our $M_{\rm h}$-$\sigma_{\rm s}$ relations are in good agreement with the simulation results in a broad halo mass range. In addition, the $M_{\rm h}$-$\sigma_{\rm s}$ relation depends weakly on $N_{\rm sat}$ and SFR. 
All of these suggest that group richness, velocity bias and merging systems can not induce the large deviations presented in these papers. 
The estimation of $\sigma_{\rm s}$ for individual groups is sensitive to interlopers and the number of group members. The large uncertainty can smear and weaken the correlation between $M_{\rm h}$ and $\sigma_{\rm s}$, as shown in previous studies. 

\subsection{The linearly combined proxies} \label{sec_combined_proxy}


The above analyses show that the $M_*$-, $\sigma_*$- and $M_{\rm AM}$-$M_{\rm h}$
relations depends on $N_{\rm sat}$ and SFR in different manners. 
It is thus interesting to check whether the combination of these proxies can provide a better proxy\citep[see][ for a relevant study]{Han2015}. In Section \ref{sec_new_proxy}, we introduce our method for generating a new proxy. 
We construct three sets of new proxies, 
$\rm Pr(M_*, M_{\rm AM}, \emph{p})$, $\rm Pr(\sigma_*, M_{\rm AM}, \emph{p})$, 
and $\rm Pr(M_*, \sigma_*, \emph{p})$. 
To check the performance of the new proxy, we defined two parameters, $d_{N_{\rm sat}}$
and $d_{\rm SFR}$ (Equation \ref{eq_d_parameter}), to quantify the dependencies of the $M_{\rm h}$-$\rm Pr$
relation on $N_{\rm sat}$ and SFR, respectively. 
A larger $d_{N_{\rm sat}}$
or $d_{\rm SFR}$ means a stronger dependence on  $N_{\rm sat}$ or SFR,
and thus a worse proxy. When the parameters are equal to one, it means that the difference is mainly dominated by the uncertainties.
Note that our tests are all performed with the B-series samples.

\begin{figure*}
    \centering
    \includegraphics[scale=0.106]{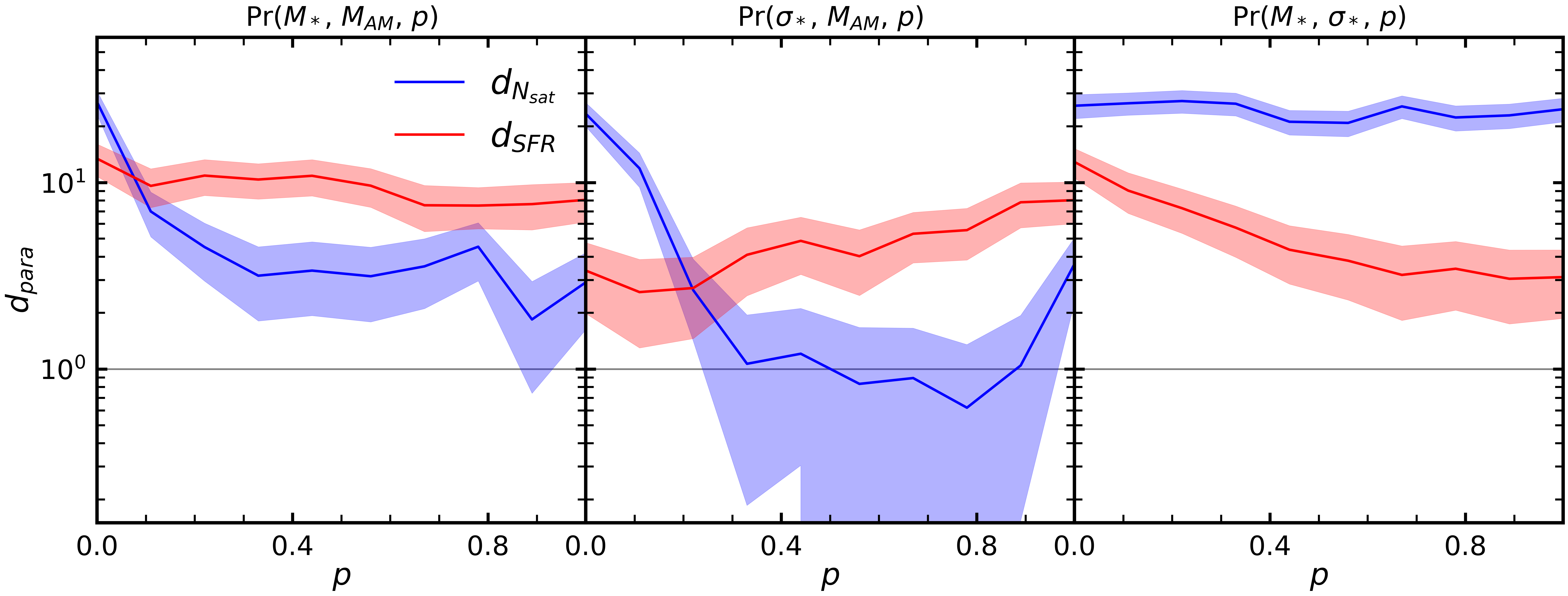}
    \caption{ The performance of the combined proxy, measured with $d_{\rm para}$(Equation \ref{eq_d_parameter}), as a function of $p$.
    The left, middle and right panels show the results for the combination between $M_*$-$M_{\rm AM}$, $\sigma_*$-$M_{\rm AM}$ and $M_*$-$\sigma_*$, respectively. In each panel, the blue and red lines correspond to the results for $d_{N_{\rm sat}}$ and $d_{\rm SFR}$, respectively. The corresponding blue and red shaded regions are the 1$\sigma$ scatter (see Section \ref{sec_new_proxy} for the detail). The gray horizontal line shows $d_{\rm para}$=1.}
    \label{fig_chis}
\end{figure*}

We first examine the combination of $M_*$ and $M_{\rm AM}$. Figure \ref{fig_chis} show
$d_{N_{\rm sat}}$ and $d_{\rm SFR}$ as functions of $p$. We can see that $d_{N_{\rm sat}}$ quickly decreases with increasing $p$ at $p<0.3$ and then remains unchanged at $d_{N_{\rm sat}}\sim3$. It means that the $M_{\rm h}$-$M_{\rm AM}$ relation is less
dependent on $N_{\rm sat}$ than the SHMR, well consistent with the above results. The parameter $d_{\rm SFR}$ is around 10 and slightly decreases
with increasing $p$. It is also consistent with the results shown above that $M_{\rm AM}$- and $M_*$-$M_{\rm h}$ relations depend on SFR in a similar manner. We thus conclude that $M_{\rm AM}$ is a better proxy than $M_*$ and that the combination of the two parameters can not improve the performance.

We then examine the combination of $\sigma_*$ and $M_*$.
The results are presented in the right panel of Figure \ref{fig_chis}. As one can see, $d_{N_{\rm sat}}$ is very large, about 25, and almost independent of $p$. It means that both $\sigma_*$- and $M_*$-$M_{\rm h}$ relations are 
strongly dependent on $N_{\rm sat}$, consistent with the results shown in Figure \ref{fig_SHMR} and Figure \ref{fig_hm_vds}. $d_{\rm SFR}$ gradually decreases with increasing $p$.
It is consistent with that the $M_{\rm h}$-$\sigma_*$ relation is less sensitive to
SFR than the SHMR. These tests suggest that $\sigma_*$ is, on average, a better
proxy than $M_*$. However, it does not mean that
$\sigma_*$ is a better proxy of halo mass than $M_*$ over any mass range.
For example, we find that $M_*$ is better than $\sigma_*$ for low mass halos (Section \ref{sec_Mh_g_sigma}). 
Note that $d_{\rm SFR}$ measures the SFR dependence averaged over a large halo mass range weighted with halo mass uncertainties.  Since the $M_{\rm h}$ uncertainties for low mass halos are much larger than those for massive halos, the contribution of low mass halos to  $d_{\rm SFR}$ is negligible.

Finally, we examine the combination of $\sigma_*$ and $M_{\rm AM}$ (the middle panel of Figure \ref{fig_chis}). $d_{N_{\rm sat}}$ decreases quickly with increasing $p$ at $p<0.3$, then almost remains unchanged around unity at $0.3<p<0.9$, and increases at  $p>0.9$. While $d_{\rm SFR}$ 
is almost constant at $p<0.3$ and then gradually increases with increasing $p$. Therefore, both $d_{N_{\rm sat}}$ and $d_{\rm SFR}$ reach the minimum
around $p=0.3$. In fact, the combined proxy $\rm Pr(\sigma_*, M_{\rm AM}, 0.3)$ has almost the smallest $d_{N_{\rm sat}}$ and $d_{\rm SFR}$ among the
three sets of combined proxies.
It means that the obtained $M_{\rm h}$-$\rm Pr(\sigma_*, M_{\rm AM}, 0.3)$ relation would has the smallest scatter among all the proxies that we have explored in this section.

As mentioned above, the two parameters $d_{N_{\rm sat}}$ and $d_{\rm SFR}$ are dominated by massive halos that have $M_{\rm h}$ measurements with high signal-to-noise ratio. They are thus unable to reflect the performance of the proxy at different mass bins, particularly at low mass. Future imaging surveys, such as
Legacy Survey of Space and Time\footnote{https://www.lsst.org/}, the Chinese Space Station Optical Survey\citep{Gong2019} and the Wide Field Survey Telescope\citep{WFST2023} can provide much better data which allows us to explore the performance of different proxies in details.

\section{Summary and discussion}\label{sec_summary}

In this paper, we use the DECaLS shear catalog to constrain the halo masses ($M_{\rm h}$) of SDSS central galaxies and investigate the scaling relations between $M_{\rm h}$ and four single halo mass proxies, including stellar mass ($M_*$), central stellar velocity dispersion ($\sigma_*$), group stellar mass ranked abundance matching halo mass ($M_{\rm AM}$) and satellite kinematics ($\sigma_{\rm s}$). We also examine the dependence of these scaling relations on galaxy internal properties, including $M_*$, $\sigma_*$ and specific star formation rate (sSFR), and external properties, including group richness($N_{\rm sat}$). These dependencies can be used to examine the performance of these halo mass proxies and understand the galaxy formation processes. We also construct a series of new halo mass proxies by linearly combining two proxies and quantify their performances. In the following, we summarize our results and briefly discuss the implications.

Overall, we find that halo mass is strongly correlated with $M_*$, $\sigma_*$, $M_{\rm AM}$ and $\sigma_{\rm s}$.
Galaxies with larger $M_*$, $\sigma_*$, $M_{\rm AM}$ and $\sigma_{\rm s}$ tend to live in more massive halos. The four obtained scaling relations are in good agreement with previous observation or simulation results. 

The $M_{\rm h}$-$M_*$ and $M_{\rm h}$-$\sigma_{\rm *}$ scaling relations are strongly dependent on $N_{\rm sat}$. At a given $M_*$ or $\sigma_{\rm *}$, the halo mass increases with $N_{\rm sat}$. 
The halo mass difference between galaxies with $N_{\rm sat}=0$ and $N_{\rm sat}\geq3$ can reach about 0.7 dex at high $M_*$ or $\sigma_{\rm *}$ ends. 
We suggest that the intrinsic dependence should be even stronger because our galaxy sample is magnitude limited. 
The halo abundance matching(HAM) technique that uses central galaxy properties as indicators of halo mass may fail to recover the halo masses of groups with large $N_{\rm sat}$. Since these groups have a large number of satellites, such a failure may cause a bias on the predicted two-point correlation function at a small scale.  
Our results further suggest that the galaxy clusters selected based on $N_{\rm sat}$ may be biased to high $M_*/M_{\rm h}$ clusters.

The $M_{\rm h}$-$M_*$ and $M_{\rm h}$-$\sigma_*$ relations are also dependent on the SFR of central galaxies. 
It means a higher baryon-to-star conversion efficiency ($M_*/M_{\rm h}/(\Omega_{\rm b}/\Omega_{\rm m})$) for star-forming galaxies than for quiescent galaxies. It may place an important constraint on AGN feedback, which is thought to suppress the conversion efficiency. 
We convert the $M_{\rm h}$-$\sigma_*$ relation into the $M_{\rm h}$-$M_{\rm BH}$ relation. At $\log M_{\rm h}/\Msun <12.0$ or $\log M_{\rm BH}/\Msun <7.4$, $M_{\rm BH}$ is weakly correlated with $M_{\rm h}$. 
It hints at a rapid black hole growth. At $\log M_{\rm h}/\Msun >12.0$ or $\log M_{\rm BH}/\Msun >7.4$, $M_{\rm BH}$ linearly correlates with $M_{\rm h}$. It indicates a transition of black hole growth. 
More studies are required to understand the underlying processes.

We investigate the correlations among $M_{\rm h}$, $M_*$ and $\sigma_*$. At small $M_*$ or $\sigma_*$ (roughly $\log M_*/\Msun <10.4$ or $\log\sigma_*<2.1$), $M_*$ is more closely related to $M_{\rm h}$ than $\sigma_*$;
In the intermediate $M_*$ or $\sigma_*$ range (roughly $10.4<\log M_*/\Msun <11.1$ or $2.1<\log\sigma_*<2.4$), $\sigma_*$ is more closely related with $M_{\rm h}$ than $M_*$; 
At large $M_*$ or $\sigma_*$ (roughly $\log M_*/\Msun >11.1$ or $\log\sigma_*>2.4$), $M_{\rm h}$ correlates with both parameters. 
The bivariate correlation suggests a complex connection among the growth of galaxy stellar mass, galaxy structure, black hole mass, and halo mass. It implies different growth patterns for galaxies at different stages. Higher quality weak lensing data are required to reexamine this issue.

Another examined halo mass proxy is $M_{\rm AM}$. Our results show that the $M_{\rm h}$-$M_{\rm AM}$ relation is close to the one-to-one relation.  Different from the SHMR and $M_{\rm h}$-$\sigma_*$ relations, the $M_{\rm h}$-$M_{\rm AM}$ relation is weakly dependent on $N_{\rm sat}$ and independent of the stellar mass of central galaxies. 
It is expected because the estimation of $M_{\rm AM}$ is based on the mass of total member galaxies, so $M_{\rm AM}$ has already included the contribution of satellites and centrals.
However, the $M_{\rm h}$-$M_{\rm AM}$ relation shows a strong dependence on SFR and $\sigma_*$.
It hints that galaxy formation processes still have an important impact on the relationship. Overall, $M_{\rm AM}$ is a better halo mass proxy than $M_*$ and $\sigma_*$.

The fourth scaling relation we investigate is the $M_{\rm h}$-$\sigma_{\rm s}$ relation. 
We use $M_*$, $\sigma_*$ and $M_{\rm AM}$ as the halo mass tracers to divide a galaxy sample into subsamples and obtain the mean $\sigma_{\rm s}$ and $M_{\rm h}$ for each subsample using stacking method. After carefully handling the interloper contamination and weighting bias, we obtain $M_{\rm h}$-$\sigma_{\rm s}$ relations in good agreement with the simulation results, in particular the hydrodynamical simulation, from $\log M_{\rm h}/\Msun <12.0$ to $\log M_{\rm h}/\Msun >14.0$. Moreover, the scaling relation shows weak dependence on
$N_{\rm sat}$ and SFR. Therefore, $\sigma_{\rm s}$ is the best proxy of halo mass among all these single proxies in consideration. It is expected because the scaling relation originates from virialized systems and is insensitive to galaxy formation processes.
However, the estimation of $\sigma_{\rm s}$ for individual groups is a serious challenge because of the interlopers and small satellite numbers. 

Finally, we construct new halo mass proxies by linearly combining two single proxies. We examine three sets of combinations, including $M_*$-$M_{\rm AM}$, $\sigma_*$-$M_{\rm AM}$ and $M_*$-$\sigma_*$ combinations. We use the dependencies of the halo mass-new proxy scaling relation on $N_{\rm sat}$ and SFR to check the performance of the new proxy. The combination of $\sigma_*$ and $M_{\rm AM}$ is better than
the other two. It is because that the $M_{\rm h}$-$M_{\rm AM}$ scaling relation is insensitive to $N_{\rm sat}$ and the $M_{\rm h}$-$\sigma_*$ scaling relation depends weakly on SFR. When $M_{\rm AM}$ contributes 30\%
and $\sigma_*$ contributes 70\% to the new proxy, the dependencies on $N_{\rm sat}$ and SFR are almost the weakest. We emphasize that our study provides a way to construct a halo mass proxy for individual groups and evaluate its performance. Future image surveys can provide much better lensing signals, which may allow us to study these proxies in detail and construct better ones.

\section*{Acknowledgements}
We thank the referee and statistics editor for their useful reports. We thank Kai Wang from Peking University for his valuable suggestions on drawing. This work is supported by CAS Project for Young Scientists in Basic Research, Grant No. YSBR-062, the National Key R\&D Program of China (grant Nos. 2018YFA0404503 and 2018YFA0404504), the National Natural Science Foundation of China (Nos. 12192224, 11733004, 11890693, 11890691, and 12073017), and the Fundamental Research Funds for the Central Universities. We acknowledge the science research grants from the China Manned Space Project with Nos. CMS-CSST-2021-A01 and CMS-CSST-2021-A03. The authors gratefully acknowledge the support of Cyrus Chun Ying Tang Foundations. 
The work is also supported by the Supercomputer Center of University of Science and Technology of China.

\bibliography{ref}{}
\bibliographystyle{aasjournal}

\begin{appendix}
\section{The construction of controlled samples}\label{sec_app}

Our purpose is to investigate the dependence of $M_{\rm h}$ on the parameter $X$ with parameter $Y$ controlled. We divide a galaxy sample within a $Y$ range into several bins according to $X$. Galaxies in these different $X$ bins have
different distributions of $Y$ parameter, although they are selected in the same $Y$ range.
To ensure that the dependence of $M_{\rm h}$ on $X$ is not affected by the $Y$ parameter,
it is necessary to construct the $Y$-controlled subsamples. 

\begin{figure*}
    \centering
    \includegraphics[scale=0.20]{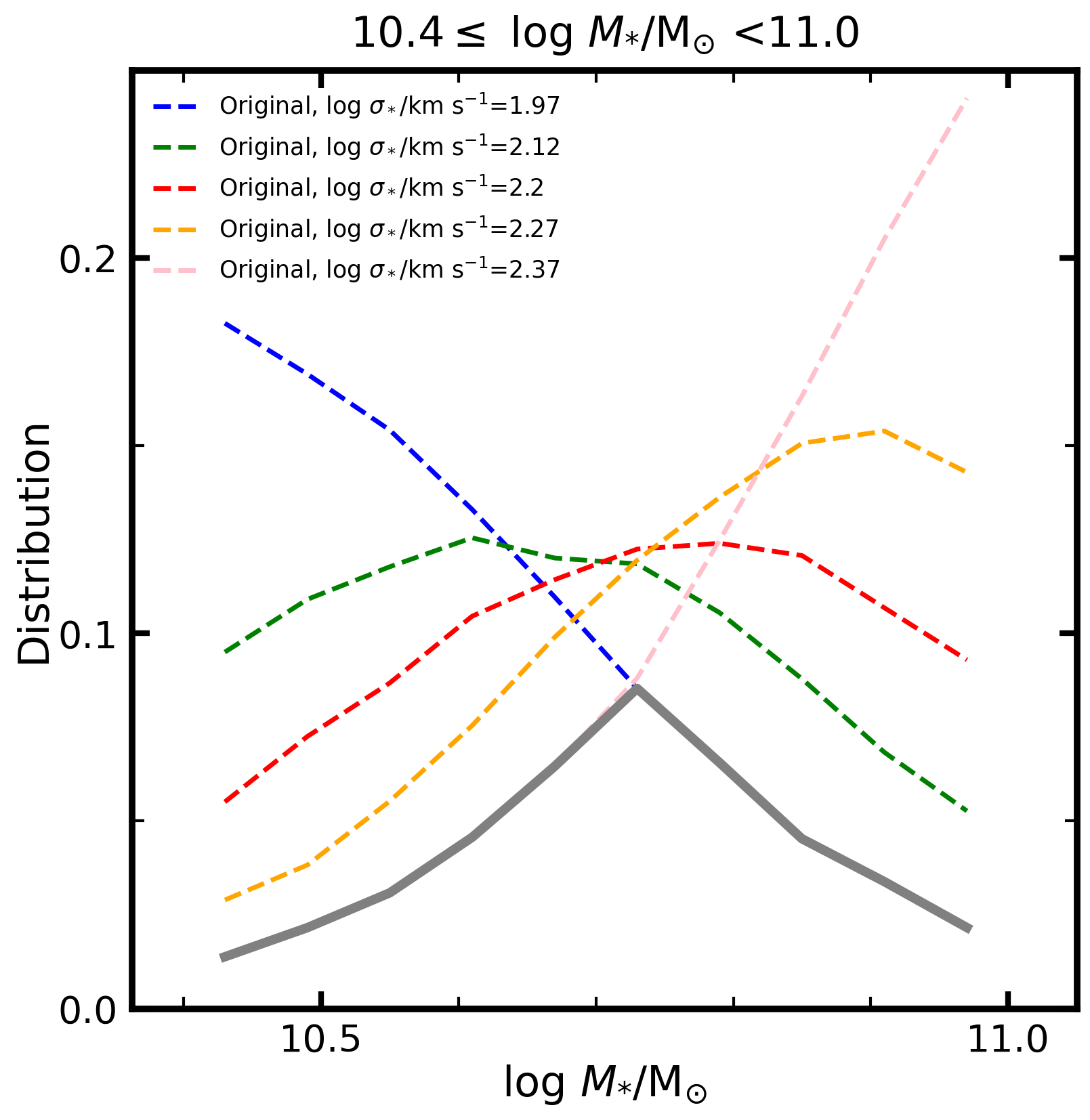}
    \caption{Stellar mass distributions of different galaxy samples in the same stellar mass bin. The color-coded dashed lines show the original $M_*$ distributions for galaxies in different $\sigma_*$ bins. These galaxies are restricted in the stellar mass range of $10.4\leq\log M_*/\Msun <11$. The thick gray line shows the lower envelope of the dashed lines. Our controlled subsamples are selected in each $\sigma_*$ bin according to the gray line.}
    \label{fig_sm_controlled}
\end{figure*}

Take the $M_*$-controlled subsamples in Section \ref{sec_Mh_g_sigma} as
an example. We want to study the correlation between $M_{\rm h}$ and $\sigma_*$ ($X$ parameter) at a given $M_*$ ($Y$ parameter).  We thus divide a galaxy sample in a given $M_*$ range into several $\sigma_*$ bins.  Figure \ref{fig_sm_controlled} shows the $M_*$ probability distributions for galaxies of $10.4\leq\log M_*/\Msun <11.0$ in five different $\sigma_*$ bins (dashed lines).
We can see that the $M_*$ distributions are still very different, even in the relatively narrow $M_*$ range. It is unsurprising because $M_*$ and $\sigma_*$ are strongly correlated. 
To construct $M_*$-controlled subsamples to eliminate the impact of $M_*$, we select galaxies in each $\sigma_*$ bin 
following the distribution indicated by the thick gray line, which
is the lower envelope of these dashed lines. 
This method ensures sufficient galaxies in every $\sigma_*$ bin.

\end{appendix}
\end{document}